\documentclass[english,showpacs,english,aps,prc,twocolumn,superscriptaddress,floatfix]{revtex4}
\usepackage[T1]{fontenc}
\usepackage[latin9]{inputenc}
\usepackage{refstyle}
\usepackage{textcomp}
\usepackage{amstext}
\usepackage{amssymb}
\usepackage{graphicx}
\usepackage{color}
\usepackage{lipsum}
\usepackage{float} 
\usepackage{tabularx}
\usepackage{mathtools}
\usepackage{ulem}		

\usepackage[percent]{overpic}

\makeatletter

\newcommand{\lyxmathsym}[1]{\ifmmode\begingroup\def\b@ld{bold}
  \text{\ifx\math@version\b@ld\bfseries\fi#1}\endgroup\else#1\fi}

\RS@ifundefined{subref}
  {\def\RSsubtxt{section~}\newref{sub}{name = \RSsubtxt}}
  {}
\RS@ifundefined{thmref}
  {\def\RSthmtxt{theorem~}\newref{thm}{name = \RSthmtxt}}
  {}
\RS@ifundefined{lemref}
  {\def\RSlemtxt{lemma~}\newref{lem}{name = \RSlemtxt}}
  {}

\@ifundefined{textcolor}{}
{%
 \definecolor{BLACK}{gray}{0}
 \definecolor{WHITE}{gray}{1}
 \definecolor{RED}{rgb}{1,0,0}
 \definecolor{GREEN}{rgb}{0,1,0}
 \definecolor{BLUE}{rgb}{0,0,1}
 \definecolor{CYAN}{cmyk}{1,0,0,0}
 \definecolor{MAGENTA}{cmyk}{0,1,0,0}
 \definecolor{YELLOW}{cmyk}{0,0,1,0}
}

\makeatother

\usepackage{babel}
\begin{document}

\title{Experimental study of the $^{40,48}$Ca+$^{40,48}$Ca reactions at 35 MeV/nucleon}

\author{Q.~Fable}
\email{quentin.fable@l2it.in2p3.fr}
\altaffiliation[Present address: ]{Laboratoire des 2 Infinis - Toulouse (L2IT-IN2P3), Universit\'e de Toulouse, CNRS, UPS, F-31062 Toulouse Cedex 9 (France)}
\affiliation{GANIL, CEA/DRF-CNRS/IN2P3, Bvd. Henri Becquerel, F-14076 Caen CEDEX, France}
\affiliation{Normandie Univ, ENSICAEN, UNICAEN, CNRS/IN2P3, LPC Caen, F-14000 Caen, France}
\affiliation{Laboratoire des 2 Infinis - Toulouse (L2IT-IN2P3), Universit\'e de Toulouse, CNRS, UPS, F-31062 Toulouse Cedex 9 (France)}

\author{A.~Chbihi}
\affiliation{GANIL, CEA/DRF-CNRS/IN2P3, Bvd. Henri Becquerel, F-14076 Caen CEDEX, France}

\author{M.~Boisjoli}
\affiliation{D\'epartement de physique, de g\'enie physique et d'optique, Universit\'e Laval, Qu\'ebec, G1V 0A6 Canada}


\author{J.D.~Frankland}
\affiliation{GANIL, CEA/DRF-CNRS/IN2P3, Bvd. Henri Becquerel, F-14076 Caen CEDEX, France}

\author{A.~Le~F\`evre}
\affiliation{GSI Helmholtzzentrum f\"{u}r Schwerionenforschung GmbH, D-64291 Darmstadt, Germany}

\author{N.~Le~Neindre}
\affiliation{Normandie Univ, ENSICAEN, UNICAEN, CNRS/IN2P3, LPC Caen, F-14000 Caen, France}

\author{P.~Marini}
\affiliation{Univ. Bordeaux, CNRS, CENBG, UMR 5797, F-33170 Gradignan, France}


\author{G.~Verde}
\affiliation{Laboratoire des 2 Infinis - Toulouse (L2IT-IN2P3), Universit\'e de Toulouse, CNRS, UPS, F-31062 Toulouse Cedex 9 (France)}
\affiliation{Istituto Nazionale di Fisica Nucleare, Sezione di Catania, 64 Via Santa Sofia, I-95123, Catania, Italy}


\author{G.~Ademard}
\affiliation{Universit\'e Paris-Saclay, CNRS/IN2P3, IJCLab, 91405 Orsay, France}

\author{L.~Bardelli}
\affiliation{Sezione INFN di Firenze, Via G. Sansone 1, I-50019 Sesto Fiorentino, Italy}

\author{C.~Bhattacharya}
\author{S.~Bhattacharya}
\affiliation{Variable Energy Cyclotron Centre, 1/AF Bidhan Nagar, Kolkata, India}

\author{E.~Bonnet}
\affiliation{SUBATECH UMR 6457, IMT Atlantique, Universit\'e de Nantes, CNRS-IN2P3, 44300 Nantes, France}

\author{B.~Borderie}
\affiliation{Universit\'e Paris-Saclay, CNRS/IN2P3, IJCLab, 91405 Orsay, France}

\author{R.~Bougault}
\affiliation{Normandie Univ, ENSICAEN, UNICAEN, CNRS/IN2P3, LPC Caen, F-14000 Caen, France}

\author{G.~Casini}
\affiliation{Sezione INFN di Firenze, Via G. Sansone 1, I-50019 Sesto Fiorentino, Italy}

\author{M.~La Commara}
\affiliation{Dipartimento di Farmacia, Universit\`{a} Federico II and INFN Napoli, Napoli, Italia}

\author{R.~Dayras} 
\affiliation{Irfu, CEA, Universi\'e Paris-Saclay, Centre de Saclay, F-91191 Gif-sur-Yvette, France}

\author{J.E.~Ducret}
\affiliation{GANIL, CEA/DRF-CNRS/IN2P3, Bvd. Henri Becquerel, F-14076 Caen CEDEX, France}

\author{F.~Farget}
\affiliation{GANIL, CEA/DRF-CNRS/IN2P3, Bvd. Henri Becquerel, F-14076 Caen CEDEX, France}

\author{E.~Galichet}
\affiliation{Universit\'e Paris-Saclay, CNRS/IN2P3, IJCLab, 91405 Orsay, France}
\affiliation{Conservatoire National des Arts et M\'etiers, F-75141 Paris Cedex 03, France}

\author{T.~G\'enard}
\affiliation{GANIL, CEA/DRF-CNRS/IN2P3, Bvd. Henri Becquerel, F-14076 Caen CEDEX, France}

\author{F.~Gramegna}
\affiliation{INFN, Laboratori Nazionali di Legnaro,  Viale dell'Universit\`{a}, 2-35020 Legnaro (PD), ITALY}

\author{D.~Gruyer}
\affiliation{Normandie Univ, ENSICAEN, UNICAEN, CNRS/IN2P3, LPC Caen, F-14000 Caen, France}


\author{M.~Henri}
\affiliation{GANIL, CEA/DRF-CNRS/IN2P3, Bvd. Henri Becquerel, F-14076 Caen CEDEX, France}

\author{S.~Kundu}
\affiliation{Variable Energy Cyclotron Centre, 1/AF Bidhan Nagar, Kolkata, India}

\author{J.~Lemari\'e}
\affiliation{GANIL, CEA/DRF-CNRS/IN2P3, Bvd. Henri Becquerel, F-14076 Caen CEDEX, France}

\author{O.~Lopez}
\affiliation{Normandie Univ, ENSICAEN, UNICAEN, CNRS/IN2P3, LPC Caen, F-14000 Caen, France}

\author{J.~\L{}ukasik}
\affiliation{H. Niewodnicza\'nski Institute of Nuclear Physics, Pl-31342 Krak\'ow, Poland}

\author{L.~Manduci}
\affiliation{\'Ecole des Applications Militaires de l'\'energie Atomique, B.P. 19, F-50115 Cherbourg, France}

\author{J.~Moisan}
\affiliation{GANIL, CEA/DRF-CNRS/IN2P3, Bvd. Henri Becquerel, F-14076 Caen CEDEX, France}

\author{G.~Mukherjee}
\affiliation{Variable Energy Cyclotron Centre, 1/AF Bidhan Nagar, Kolkata, India}

\author{P.~Napolitani}
\affiliation{Universit\'e Paris-Saclay, CNRS/IN2P3, IJCLab, 91405 Orsay, France}

\author{A.~Olmi}
\affiliation{Sezione INFN di Firenze, Via G. Sansone 1, I-50019 Sesto Fiorentino, Italy}

\author{M.~P\^arlog}
\affiliation{Normandie Univ, ENSICAEN, UNICAEN, CNRS/IN2P3, LPC Caen, F-14000 Caen, France}
\affiliation{National Institute for Physics and Nuclear Engineering, RO-077125 Bucharest-M\u{a}gurele, Romania}

\author{S.~Piantelli}
\affiliation{Sezione INFN di Firenze, Via G. Sansone 1, I-50019 Sesto Fiorentino, Italy}

\author{G.~Poggi}
\affiliation{Sezione INFN di Firenze, Via G. Sansone 1, I-50019 Sesto Fiorentino, Italy}

\author{A.~Rebillard-Souli\'e}
\affiliation{Normandie Univ, ENSICAEN, UNICAEN, CNRS/IN2P3, LPC Caen, F-14000 Caen, France}

\author{R. Roy}
\affiliation{D\'epartement de physique, de g\'enie physique et d'optique, Universit\'e Laval, Qu\'ebec, G1V 0A6 Canada}

\author{B.~Sorgunlu}
\affiliation{GANIL, CEA/DRF-CNRS/IN2P3, Bvd. Henri Becquerel, F-14076 Caen CEDEX, France}

\author{S.~Velardita}
\affiliation{Institut f\"{u}r Kernphysik, Technische Universit\"{a}t Darmstadt, D-64289 Darmstadt, Germany.}

\author{E.~Vient}
\affiliation{Normandie Univ, ENSICAEN, UNICAEN, CNRS/IN2P3, LPC Caen, F-14000 Caen, France}

\author{M.~Vigilante}
\affiliation{Dipartimento di Fisica, Universit\`a degli Studi di Napoli FEDERICO II, I-80126 Napoli, Italy}
\affiliation{Istituto Nazionale di Fisica Nucleare, Sezione di Napoli, Complesso Universitario di Monte S. Angelo, Via Cintia Edificio 6, I-80126 Napoli, Italy}
\author{J.P.~Wieleczko}
\affiliation{GANIL, CEA/DRF-CNRS/IN2P3, Bvd. Henri Becquerel, F-14076 Caen CEDEX, France}

\collaboration{INDRA collaboration}\noaffiliation

\begin{abstract}
In this article we investigate $^{40,48}$Ca+$^{40,48}$Ca peripheral and semi-peripheral reactions at 35 MeV/nucleon. 
Data were obtained using the unique coupling of the VAMOS high acceptance spectrometer and the INDRA charged particle multidetector.
The spectrometer allowed high resolution measurement of charge, mass and velocity of the cold projectile-like fragment (PLF), while the INDRA detector recorded coincident charged particles with nearly $4\pi$ acceptance.
The measured isotopic composition of the PLF identified in VAMOS and the average light charged particle (LCP) multiplicities are promising observables to study the isospin diffusion.
The detection of the PLF in coincidence with LCP allows the reconstruction of the mass, charge and excitation energy of the associated initial quasi-projectile nuclei (QP), as well as the extraction of apparent temperatures.
We investigate the suitability of the isoscaling method with the PLF and the experimental reconstructed QP.
The extracted $\alpha$ and $\Delta$ isoscaling parameters present a dependence on the considered system combination that could justify their use as a surrogate for isospin asymmetry in isospin transport studies.
The reconstruction of the QP allows to observe an evolution of the $\alpha/\Delta$ with the size of the QP, the latter being consistent with a strong surface contribution to the symmetry energy term in finite nuclei.  
This leads to the conclusion that the reconstruction of the primary source is mandatory for the study of the symmetry energy term based on the isoscaling method for such reactions.

\end{abstract}
\date{\today}
\maketitle

\section{Introduction}\label{sec_intro}

The nuclear equation of state (EOS) is a fundamental property of nuclear matter, describing the relationship between energy, temperature, density and neutron-to-proton asymmetry of the system. It plays an important role in the supernova phenomenon \cite{Fischer2014}, the formation, cooling and structure of neutron stars \cite{Lattimer2014, Raduta2014, Gandolfi2014, huth2021constraining}, and in the mergers of compact astrophysical objects such as neutron stars and black holes \citep{LATTIMER2000121, GW170817, Abott_PhysRevLett_121_161101}.

The EOS for symmetric nuclear matter has been very extensively explored and constraints on its stiffness have been largely discussed \citep{danielewicz2002:symmetryEnergy, REISDORF20121, WANG2018207, LEFEVRE2016112}. 
Nonetheless, the symmetry energy defined as the isospin-dependent part of the EOS of asymmetric nuclear matter is still less known, in particular far from the saturation density $\rho_0$.

Several recent experimental and theoretical studies are devoted to the search for the density dependence of the nuclear symmetry energy term, see for instance the topical issue on nuclear symmetry energy \cite{Li:2014oda} and \citep{bao2002:symmetryEnergy, chen2003:symmetryEnergy, Tsang2001:symmetryEnergy, Tsang2004:isospindiffusion,
souliotis2003:isotopicScaling, shetty2007:densityDependenceEsym, shetty2007_bis:densityDependence, ono2004:SymmEnergy, Li2008113:Observables, Lefevre_2018, Russotto2016:PhysRevC.94.034608}.
Furthermore, the recent availability of accelerator facilities capable of producing both stable and radioactive beams over a wide range of neutron-to-proton asymmetries has stimulated further experimental programs devoted to exploring the EOS of asymmetric nuclear matter.

These efforts are further stimulated by the fact that the symmetry energy, representing the energy cost of converting all protons in symmetric matter into neutrons (at fixed temperature and density), determines several properties of the inner crust of neutron stars \citep{lattimer2001:NeutronStar,link1999:PulsarConstraintsNS, Steiner2005:IsospinNS, Jmargueron1, Jmargueron2} as well as the nuclear masses \cite{Pearson2014}, and the features of exotic nuclear systems, like neutron halos, where regions of very neutron-rich nuclear matter at low density are expected.
The existence of neutron skins at the surface of n-rich nuclei is also expected to be sensitive to the symmetry energy (see the review article \citep{Thiel_2019} and ref. \citep{Pruitt_Charity_SOB_2020_nskin, JunXuPhysRevC.102.044316}), as well as pygmy and giant resonances \citep{Colo2014, PhysRevC_92_024316, PhysRevC_95_034324, PhysRevC_101_064314}.

Heavy-ion collisions (HIC) allow to probe the nuclear EOS under laboratory controlled conditions, over a wide range of density and energy, depending on the incident beam energy, the size of the colliding systems and the impact parameter of the collisions.  
A variety of observables measured in HIC, mostly related to the isotopic composition of the fragments produced in the decay of the formed excited nuclear systems, are expected to be sensitive to the nuclear EOS. 
Among these extensively investigated observables, we can cite isobaric yield ratios \cite{PhysRevC_89_011001, MALLIK2013282}, collective flow \cite{Russotto2014} and isoscaling \cite{Tsang2001:symmetryEnergy, souliotis2003:isotopicScaling, SOULIOTIS200435, lefevre2005_PhysRevLett_94_162701, PhysRevC_79_061602, PhysRevC_98_044602, Marini_PhysRevC_85_034617, MALLIK2013282}.
Isospin diffusion between two nuclei with different isospin asymmetry \cite{Shi2003:isospinDiffusionTheor, Tsang2004:isospindiffusion, Sun2010:isospinDiffusion, PhysRevC_79_064615, Camaiani_PhysRevC_103_014605, Piantelli_PhysRevC_103_014603}, along with neck dynamics and emissions \cite{DeFilippo2014}, is another phenomenon allowing to probe the density dependence of the symmetry energy.
Indeed, in the framework of the Stochastic Mean Field (SMF) model, M. Colonna \textit{et ~al.} proposed the study of the neutron-to-proton ratio of the fragments produced in HIC, in given kinetic energy intervals \cite{Colonna2008:isospinDistillation}. 
In that work, a clear sensitivity to the parametrization of the symmetry energy was observed.

Constraints on the symmetry energy of finite nuclei around saturation density mainly come from fitting the Bethe-Weizs\"acker semi-empirical mass formula to the binding energies of ground-state nuclei.
Historically, it was pointed out that the symmetry energy should be mass-dependent and expressed as the sum of a volume and a surface contribution proportional to $(N-Z)^2/A$ \cite{Bethe_1971}.

Myers and Swiatecki estimated the surface to volume ratio to be $1.1838$ \cite{Myers19661}, nonetheless the volume and surface contributions have been little investigated until now. 
More recently, as pointed out by Danielewicz \textit{et.~al}, the ratio of the two components is closely related to the neutron-skin thickness \cite{DANIELEWICZ_NPA922}.
It is also expected to be sensitive to the temperature for finite nuclei as calculations showed that the surface symmetry energy term is more sensitive to temperature than the volume term \cite{Agrawal_EPJA_2014}.

This work focuses on the experimental results of peripheral and semi-peripheral collisions in the Fermi energy domain ($\approx20-100$ MeV/nucleon bombarding energies). 
According to dynamical transport models, such collisions exhibit mainly a binary character: projectile and target nuclei interact by exchanging nucleons before re-separating into a quasi-projectile (QP) and a quasi-target (QT), with kinematic properties respectively close to the projectile and the target \cite{BARAN2004329, PhysRevC_72_064620, PhysRevC_79_064615}. 
As the QP and the QT may well be moderately deformed and excited, they undergo secondary decays by emitting light particles and gamma rays.  
Thus, QP and QT remnants, respectively called projectile-like fragment (PLF) and target-like fragment (TLF), and light particles are expected to be produced in the exit channel of the reaction. In addition to those main sources of fragments, one needs to take into account the existence of a third one at mid-rapidity. 
Indeed, this transient neck-like structure is expected to be formed at intermediate velocities between the projectile and the target. 
In early works of the INDRA collaboration, a quantitative evaluation of the contribution of the mid-rapidity source has been estimated, as a function of incident energy and impact parameter \cite{I10-Luk97, I17-Pla99}. 
As a conclusion of these works, the onset of this mid-rapidity emission starts at Fermi energy and increases with incident energy.

The observables sensitive to the nuclear EOS can furthermore be significantly modified by secondary decays, leading to a possible distortion of the estimated symmetry energy coefficient. 
Indeed, depending on the chosen observable, the effect of particle emissions is not straightforward. 
As an example, a theoretical study using the Lattice Gas Model, which implicitly accounts for secondary decays, suggested the isoscaling of the PLF to be a promising observable \cite{Lehaut2009:Isoscaling}. 
In contrast, various experimental and theoretical studies have emphasized the fact that secondary decays must be taken into account to obtain meaningful and comparable results \cite{lefevre2005_PhysRevLett_94_162701, Marini2013:secondaryDecay,I39-Hud03,I11-Mar98, PhysRevC_79_061602, PhysRevC_79_064615, CHAUDHURI2011190}.
Furthermore, recent investigations on peripheral collisions of Ca isotopes with the AMD transport model followed by different evaporation models, have highlighted that both de-excitation of primary fragments and the fast dynamical emissions can affect the neutron-proton equilibration estimated via isospin transport ratio \cite{Camaiani_PhysRevC_102}.
Thus, an investigation of the symmetry energy with the aforementioned methods requires the measurement of the isotopic distributions of fragments, as well as the detection of light particles emitted in coincidence.

This work is organized as follows.
Sections \ref{sec_ExpSetup} and \ref{sec_simus} present the experimental setup and the simulation codes, respectively. 
Sec.\ref{sec_exp_results} focuses on the general experimental results for the fragment measured in VAMOS and the light charged particles detected in coincidence with INDRA.
Sec.\ref{sec_csym} is dedicated to the source reconstruction and the study of the isoscaling method applied to the data.
Finally, conclusions are reported on Sec.\ref{sec:Conclusion}.

\section{Experimental setup\label{sec_ExpSetup}}

\begin{figure}[ht]
\centering
\includegraphics[scale=0.65]{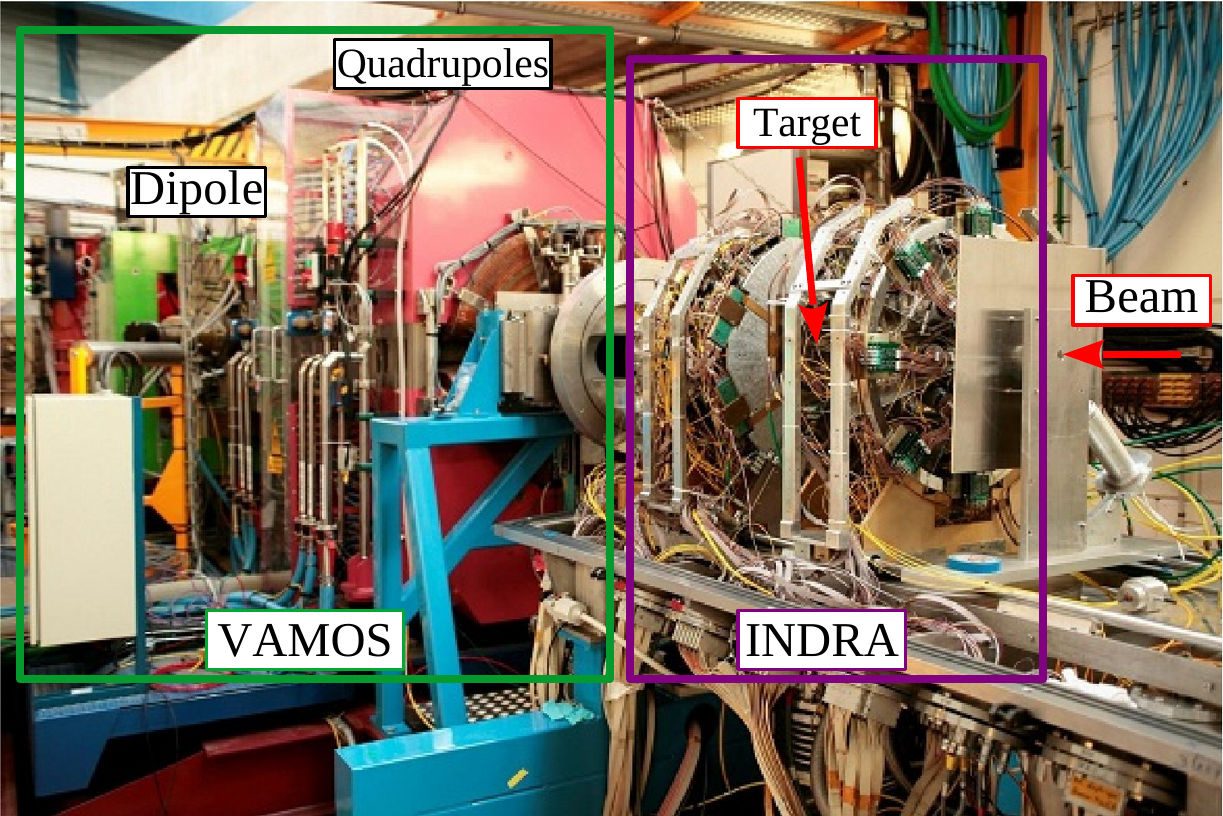}
\caption{(Color online) Picture of the experimental setup of the INDRA-VAMOS coupling.}
\label{fig_IndraVamos}
\end{figure}

The experiment was performed at the GANIL facility, where beams of $^{40,48}$Ca at 35 MeV/nucleon impinged on self-supporting $1.0$ $mg/cm^{2}$ $^{40}$Ca or $1.5$ $mg/cm^{2}$ $^{48}$Ca targets placed inside the INDRA vacuum chamber. The typical beam intensity was around $5.10^{7}$ pps. 
The detection system consisted of the coupling of the $4\pi$ charged particle array INDRA \cite{I3-Pou95,I5-Pou96} and the VAMOS spectrometer \cite{Pullanhiotan2008343:VAMOS}. Table \ref{tab_physics_systems_e503} gives a summary of the characteristics of the studied systems and Fig.\ref{fig_IndraVamos} shows a picture of the experimental setup.

\begin{table}[ht]
\centering
	\begin{tabular}{c c c c c c c}
	\hline
    Beam & $E_{inc}$ & $B\rho _0$ & $v_{lab}$ & Target  & $I_{sys}$ & $\theta _{gr}$\\
         & $(MeV/nuc)$  & $(T\,m)$    & $(cm/ns)$ &         &           & $(deg)$\\     
    \hline 
    $^{40}Ca^{18+}$ & 34.81 & 1.904 & 7.978 & $^{40}Ca$ & 1.0 & 2.35\\
    $^{40}Ca^{18+}$ & 34.81 & 1.904 & 7.978 & $^{48}Ca$ & 1.2 & 2.29\\ 
    $^{48}Ca^{20+}$ & 34.83 & 2.061 & 7.980 & $^{40}Ca$ & 1.2 & 1.91\\  
    $^{48}Ca^{20+}$ & 34.83 & 2.061 & 7.980 & $^{48}Ca$ & 1.4 & 1.86\\            
	\hline	
	\end{tabular}
\caption{Characteristics of the studied systems with: $E_{inc}$, $B\rho _0$ and $v_{lab}$ respectively the beam incident energy, magnetic rigidity and velocity in the laboratory frame, $\theta _{gr}$ the grazing angle and $I_{sys}$ the initial neutron-to-proton ratio of the total system.}
\label{tab_physics_systems_e503}	
\end{table}
 
Concerning the charged particle multidetector array INDRA, the detection telescopes are arranged in rings centered around the beam axis. In this experiment, INDRA covered polar angles from $7^{\circ}$ to $176^{\circ}$.  
Rings 1 to 3 were removed to allow the mechanical coupling with VAMOS in the forward direction. 
Rings 4 to 9 ($7^{\circ}-45^{\circ}$) consisted each of 24 three-layer detection telescopes : a gas-ionization chamber operated with C$_{3}$F$_{8}$ gas at low pressure, a 300 or 150 $\mu m$ silicon wafer and a CsI(Tl) scintillator (14 to 10 $cm$ thick) read by a photomultiplier tube. 
Rings 10 to 17 ($45^{\circ}-176^{\circ}$) included 24, 16 or 8 two-layer telescopes: a gas-ionization chamber and a CsI(Tl) scintillator of 8, 6 or 5 $cm$ thickness. 
Fragment identification thresholds are about $0.5$ and $1.5$ MeV per nucleon for the lightest ($Z \lesssim 10$) and the heaviest fragments, respectively. INDRA allows charge and isotope identification up to Be-B and only charge identification for heavier fragments. A detailed description of the INDRA detector and its electronics can be found in \cite{I3-Pou95,I5-Pou96}.

The VAMOS spectrometer is composed of two large magnetic quadrupoles focusing the incoming ions in the vertical and horizontal planes and a large magnetic dipole, which bends the trajectory of the ions.

\begin{figure}[ht]
\centering
\includegraphics[scale=0.43]{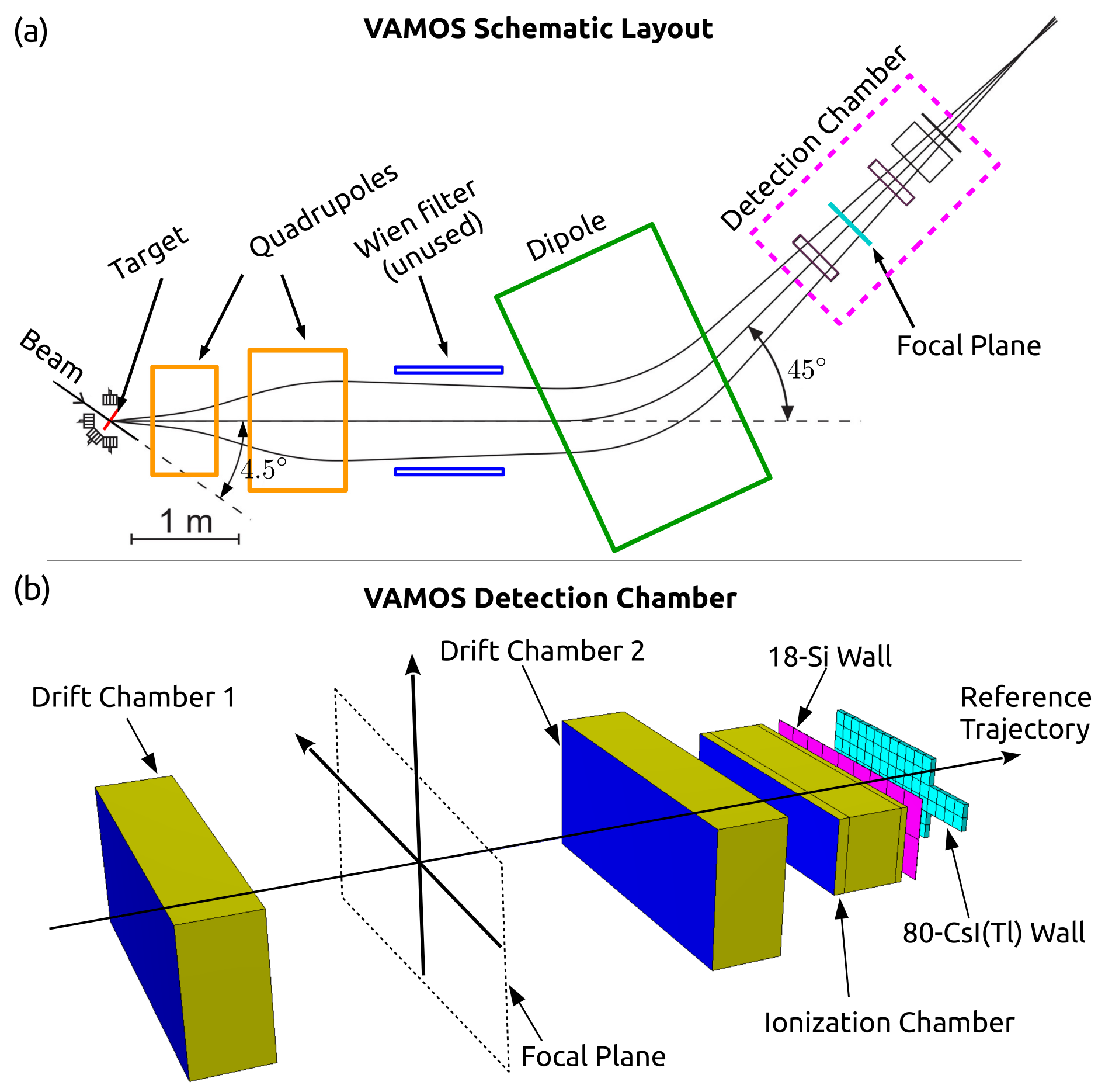}
\caption{(Color online) 
(a) VAMOS schematic layout. The spectrometer was rotated at $4.5^{\circ}$ with respect to the beam axis for the present experiment. (b) Global 3-dimensional view of VAMOS detection chamber. The reference frame corresponds to the reference trajectory. Adapted from \cite{Pullanhiotan2008343:VAMOS}.}
\label{fig_vamosdetection}
\end{figure}

A representation of the VAMOS optical line is given in Fig.\ref{fig_vamosdetection}(a). In the present setup, the spectrometer was rotated at $4.5^{\circ}$ with respect to the beam axis, so as to cover the forward polar angles from $2.56^{\circ}$ to $6.50^{\circ}$, thus favoring the detection of a fragment emitted slightly above the grazing angles of the studied reactions. The momentum acceptance was about $\pm5\%$, and the focal plane was located $9$ $m$ downstream, giving a large enough Time of Flight (ToF) base to allow a mass resolution of about $\Delta A/A \sim 1/165$ for the isotopes produced in the collisions. Further details about the mass identification achieved with the spectrometer are given in Appendix \ref{app1_VamosID}.

A three-dimensional view of the spectrometer detection chamber, located upstream and downstream of the focal plane (FP), is presented in Fig.\ref{fig_vamosdetection}(b).
The VAMOS detection setup of the experiment included two position-sensitive drift chambers used to determine the trajectories of the reaction products at the focal plane, followed by a sandwich of detectors : a 7-modules ionization chamber, a 500 $\mu m$ thick Si-wall (18 independent modules) and a $1$ $cm$ thick CsI(Tl)-wall (80 independent modules), allowing the measurements of the ToF, energy loss ($\Delta E$) and energy ($E$) parameters. The identification and reconstruction procedures of the fragments detected in the VAMOS focal plane are described in Appendix \ref{app1_VamosID}.
Around twelve magnetic rigidity ($B\rho_0$) settings, from $0.661$ to $2.220$ $T\,m$, were used for each system to cover the full velocity range of the fragments. A description of the event normalization procedure is given in Appendix \ref{app2_normalization}.

At least one hit on the VAMOS silicon wall was required for each event to be acquired, thus selecting mainly semi-peripheral and peripheral collisions. Other trigger configurations, allowing to select more central collisions, were also set but will not be discussed in the present work.
It should be noted that only multiplicity ``$1"$ events in the VAMOS Si-wall are considered. 
This offline selection was applied to make sure that the positions measured in the drift chambers are correct and to avoid events with ambiguous trajectory reconstruction. 
The elastic-like events (corresponding to events with no hit in INDRA and a fragment identical to the projectile in VAMOS) were also removed offline.

The INDRA-VAMOS coupling allowed, for the first time, the measurement of the isotopic yield of the whole charge range produced in the reactions in the angular range from $2.56^{\circ}$ to $6.50^{\circ}$, in coincidence with all other charged products in the angular range $7^{\circ}$ to $176^{\circ}$.

\section{Simulation codes}\label{sec_simus}

In order to better apprehend the experimental results, simulations of $^{40,48}$Ca$+^{40,48}$Ca collisions at 35 MeV/nucleon have been investigated in the framework of the Antisymmetrized Molecular Dynamics (AMD) model coupled to the statistical decay code GEMINI++ as afterburner.

\subsection{AMD and GEMINI++}\label{subsec_models}

The microscopic transport model AMD \cite{ONO2004501} was chosen to describe the dynamical evolution of the collisions. 
The input impact parameter of the simulation followed a triangular distribution from $b=0$ to the grazing value $b_{max} \simeq 8.5$ $fm$. 
Collisions were followed with a time-step of $0.75$ $fm/c$ up to $t_{lim}=300$ $fm/c$ (sufficiently long time limit at which the fragment multiplicities are considered as constant and the dynamical phase is supposed to be over).
Around $10^6$ events were produced for each system.

Our goal was not to investigate the stiffness of the nuclear EOS from the model but to compare its general predictions with the data and apprehend the effect of the experimental filter.  

The AMD primary events produced at $300$ $fm/c$ have then been used as input for the evaporation model GEMINI++ \cite{Charity2006PhysRevLett.97.162503, charity98}. For each primary event, $50$ secondary events were produced with GEMINI++.

\subsection{Filter and event sorting}\label{subsec_filter_simus}

The simulated events were filtered with a software replica of the experimental setup so as to allow direct comparisons between the model predictions and the experimental data. In addition, the same offline conditions as the experiment were applied to the filtered events.

Concerning VAMOS, cuts were applied according to the experimental polar and azimuthal angular distributions in the laboratory frame (obtained from trajectory reconstruction, see Appendix \ref{app11_event_recon} for more details). The energy thresholds of silicon and CsI detectors were also considered.

Concerning INDRA, the detection of the events was simulated within the \textsc{KaliVeda} framework \cite{KaliVeda}, with a complete description of the detector (including geometrical coverage, detector resolutions and identification thresholds).

\begin{figure}[ht]
\centering
\includegraphics[scale=0.42]{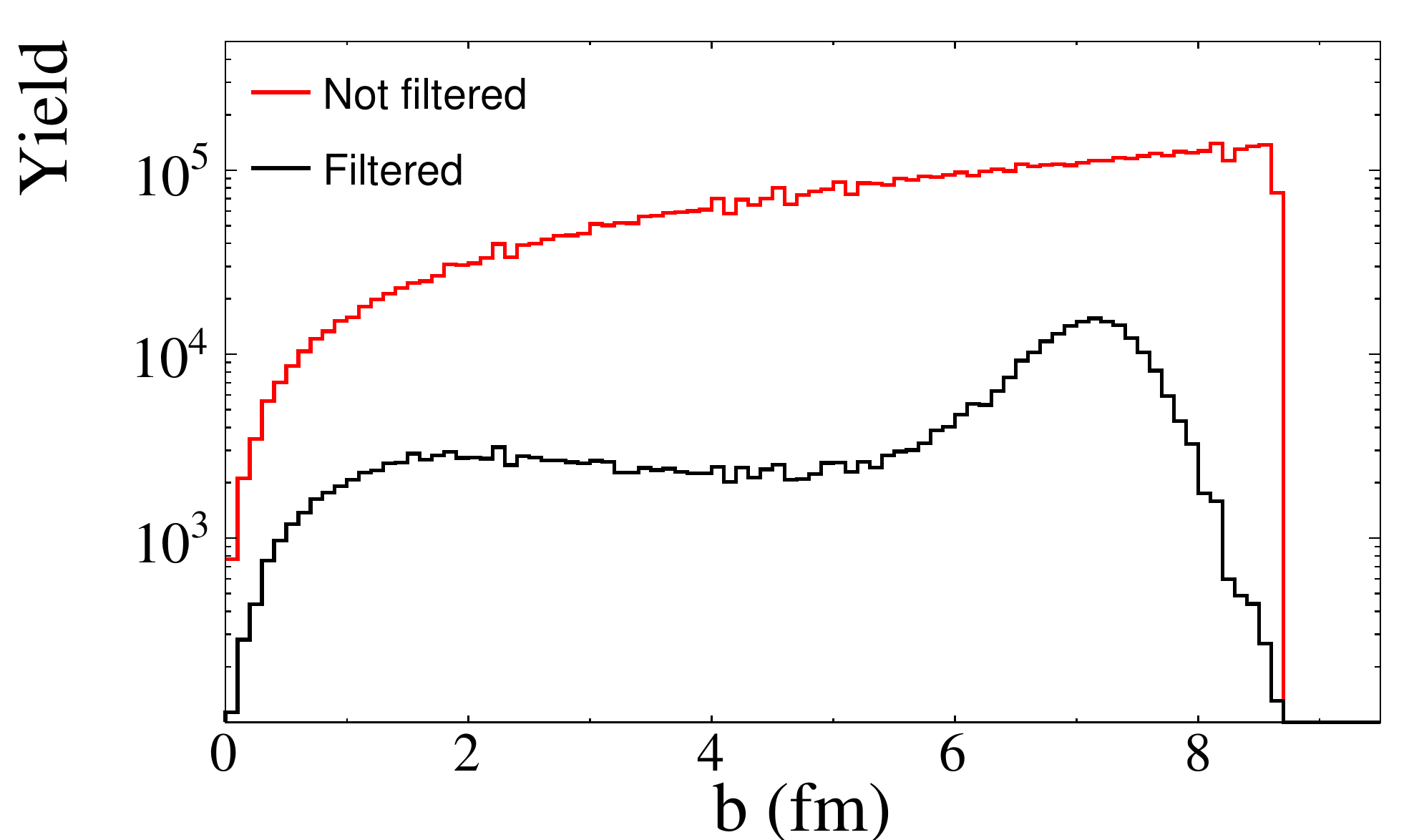}
\caption{(Color online) Effect of the VAMOS filter on the impact parameter distribution from AMD for the $^{48}$Ca$+^{48}$Ca system.}
\label{fig_impactparam}
\end{figure}

It is important to note that the spectrometer trigger condition and the associated angular acceptance filter discard most of the events ($\approx  90\%$ of the whole statistics). 
Fig.\ref{fig_impactparam} shows the distribution of the impact parameter from AMD before and after the filter for $^{48}$Ca$+^{48}$Ca collisions. 
We clearly observe the effect of the VAMOS angular cuts that favor the detection of semi-peripheral and peripheral events ($ 6 \lesssim b \lesssim 8$ $fm$).
Similar results are obtained for all the systems under study.

%

\section{Experimental results}\label{sec_exp_results}

In this section we present an overview of the reaction products, before studying the characteristics of the fragments detected in VAMOS and the charged products detected in INDRA in coincidence. 
It should be noted that a large effort has been made for the trajectory reconstruction with VAMOS and to take into account the acceptance of the spectrometer for a normalization of the statistical weight, on a event-by-event basis. More details are given in Appendices \ref{app1_VamosID} and \ref{app2_normalization}.

\subsection{Overview of the reaction products}\label{subsec_overview_exp}

\begin{figure}[ht]
\centering
\includegraphics[scale=0.45]{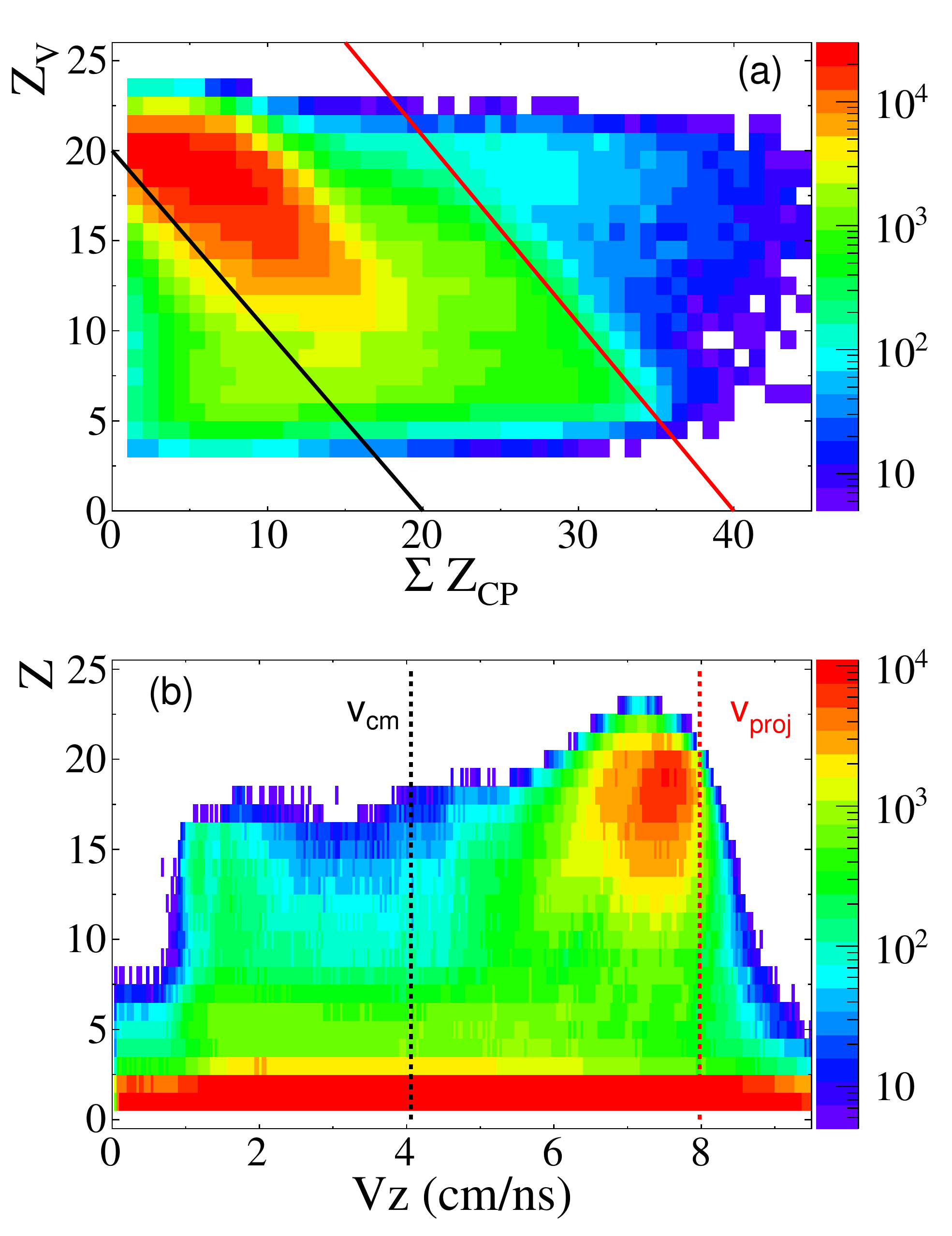}
\caption{(Color online) 
(a) Atomic number of the fragment identified in VAMOS ($Z_{V}$) as function of the sum of atomic number of the particles collected with INDRA for the $^{48}$Ca$+^{48}$Ca system. The lines indicate a complete charge detection of the total system, $Z_{proj}+Z_{target}$ = 40 (red line) and charge conservation for projectile (black line).
(b) Atomic number as a function of the parallel velocity in the laboratory frame of the nuclei identified in VAMOS and INDRA, the black and red dashed vertical lines indicate the reaction center of mass and the projectile velocities respectively.}
\label{fig_overview_exp}
\end{figure}

An overview of the events recorded for the $^{48}$Ca+$^{48}$Ca reaction is shown, as an example, in Fig.\ref{fig_overview_exp}. 
The atomic number of the fragment identified in VAMOS as a function of the sum of the atomic numbers of the charged products (CP) detected in coincidence in INDRA is presented in Fig.\ref{fig_overview_exp}(a).
The black and red lines indicate a total detected charge ($Z_{tot} =  Z_{V} + \sum Z_{CP}$) equal to the charge of the projectile ($Z_{proj}=20$) and the system ($Z_{sys}=40$), respectively. 
We observe that most of the recorded events are located between these lines, indicating a good detection efficiency for the decay products of the forward-emitted excited QP nuclei, and correct correlation between the two devices. 
For a fraction of events a complete detection of all the reaction products is even achieved. 
One can also observe the low background exceeding the total charge of the system. These events are attributed to pile-up.

Fig.\ref{fig_overview_exp}(b) depicts the charge of the nuclei identified in INDRA and VAMOS as a function of their parallel velocity in the laboratory frame. 
We observe two main components from either side of the center of mass velocity, with a third region of light charged particles (LCP, $Z<3)$ and intermediate mass fragments (IMF, $3 \leq Z \leq 10$) spreading over the entire velocity domain.
The right-most component, concentrated in a region of charge and velocity close to the projectile ($ 10 < Z < 20$ and $6<V_{z}<8$ $cm/ns$), corresponds to the fragments identified in VAMOS and assigned to the PLF, while the left-most component correspond to a TLF occasionally identified in INDRA at backward angles. 

The aforementioned observations indicate that the fragments detected in VAMOS are compatible with dissipative binary collisions and are mostly the products of the QP decay resulting from peripheral to semi-peripheral collisions. 

\begin{figure}[ht]
\centering
\includegraphics[scale=0.45]{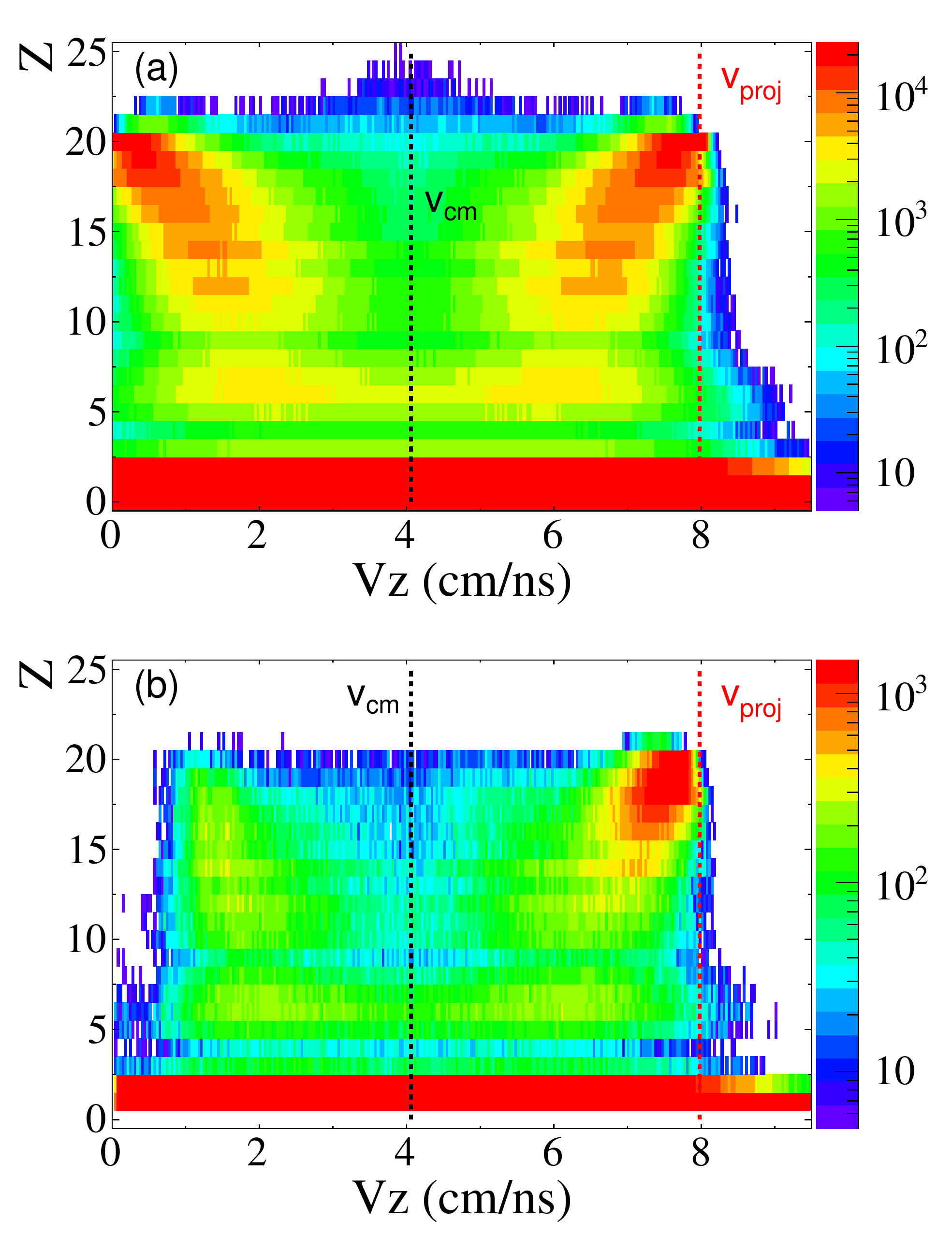}
\caption{(Color online) Atomic number $Z$ as a function of the parallel velocity (in the laboratory frame) of the secondary fragments from AMD after GEMINI++ de-excitation, for the $^{48}$Ca$+^{48}$Ca system.
(a) Without experimental filter.
(b) With experimental filter.}
\label{fig_overview_amd}
\end{figure}

Fig.\ref{fig_overview_amd} shows a comparison between all the reaction products from AMD after GEMINI++ secondary events, with and without the application of the experimental filter. 
We clearly observe the effect of VAMOS angular acceptance, more specifically the induced cut in polar angle, which drastically reduces the measured yields.
Finally, an overall agreement between the filtered simulated events and the data is observed when comparing Fig.\ref{fig_overview_exp}(b) and Fig.\ref{fig_overview_amd}(b).

\subsection{Isotopic distributions of the PLF} \label{subsec_isotopic_distri_PLF}

\begin{figure}
\includegraphics[scale=0.42]{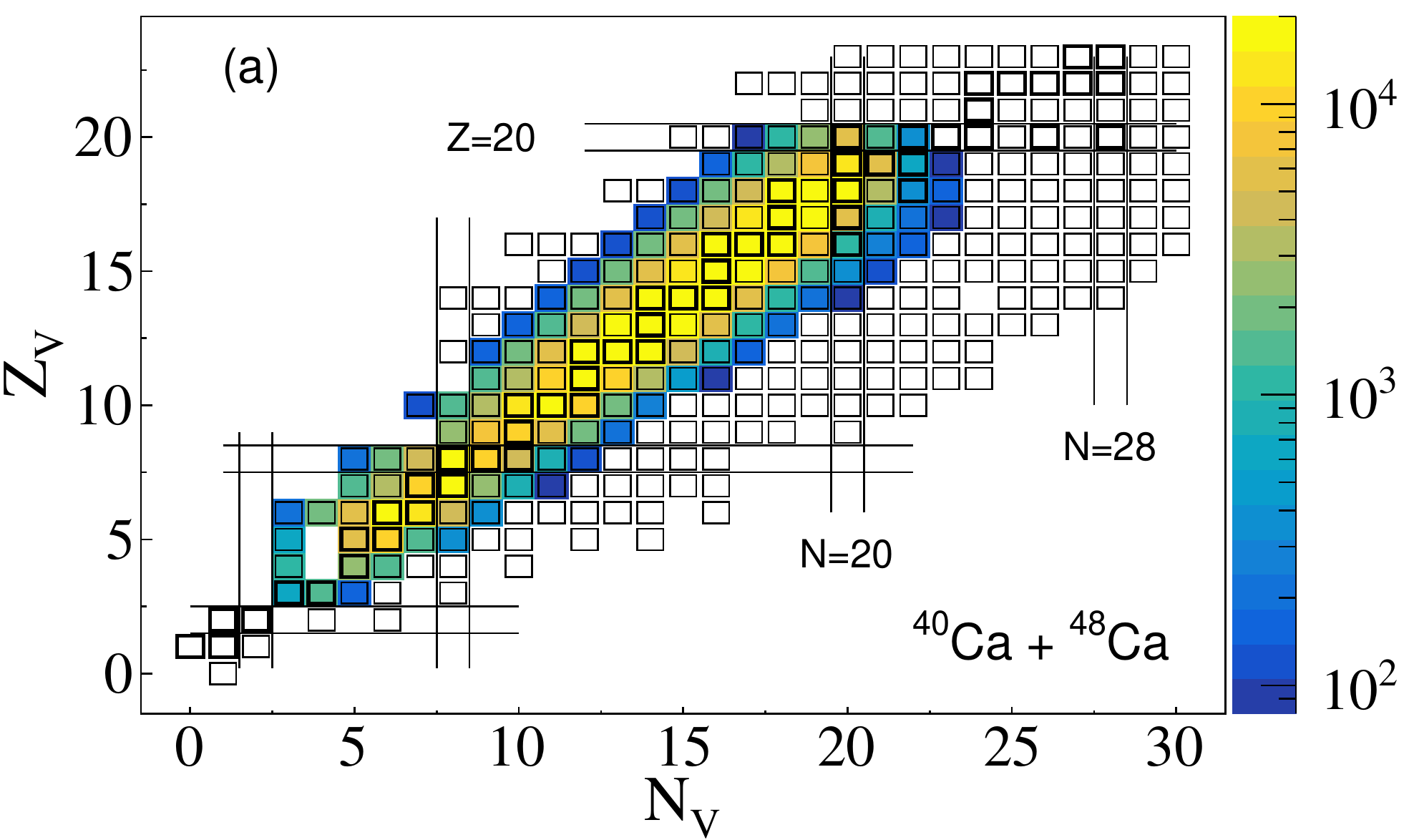}
\includegraphics[scale=0.42]{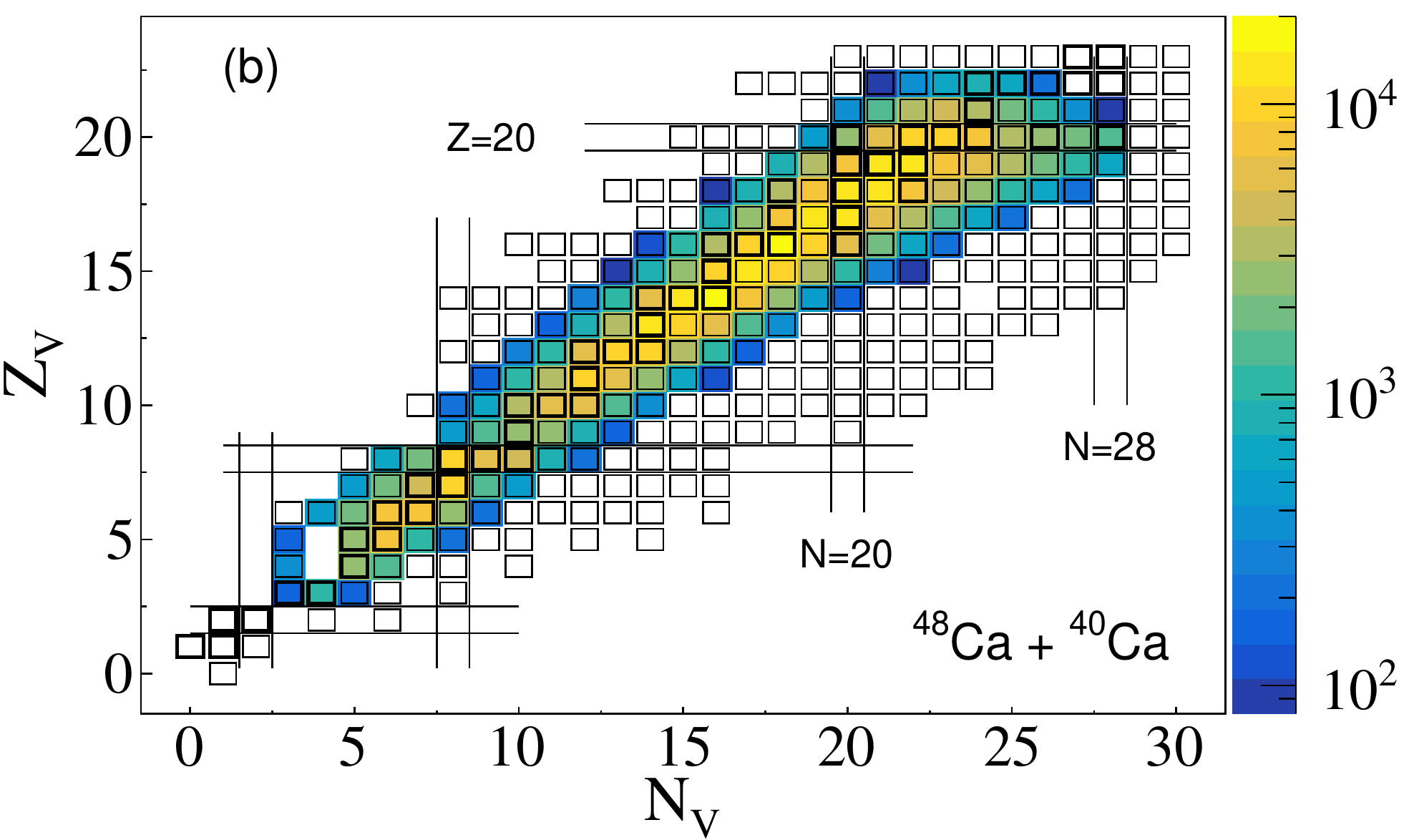} 
\caption{(Color online) Chart of the nuclides identified in VAMOS. 
(a) $^{40}$Ca+$^{48}$Ca reaction.
(b) $^{48}$Ca+$^{40}$Ca reaction.}
\label{fig_nuchar}
\end{figure}

The experimental isotopic distributions of the fragments identified in VAMOS are shown in Fig.\ref{fig_nuchar} for the $^{40}$Ca$+^{48}$Ca and $^{48}$Ca$+^{40}$Ca asymmetrical systems, in the form of two charts of nuclides. 
We observe that a broad range of isotopes is produced, with atomic numbers ranging from $3$ to $22$. 
Furthermore, a more spread-out distribution of neutron-rich nuclei for the $^{48}$Ca$+^{40}$Ca system is observed compared to the $^{40}$Ca$+^{48}$Ca.
This indicates that the memory of the projectile neutron richness is partially preserved in the recorded events, while being affected by a process of secondary evaporations.

\begin{figure}
\includegraphics[scale=0.42]{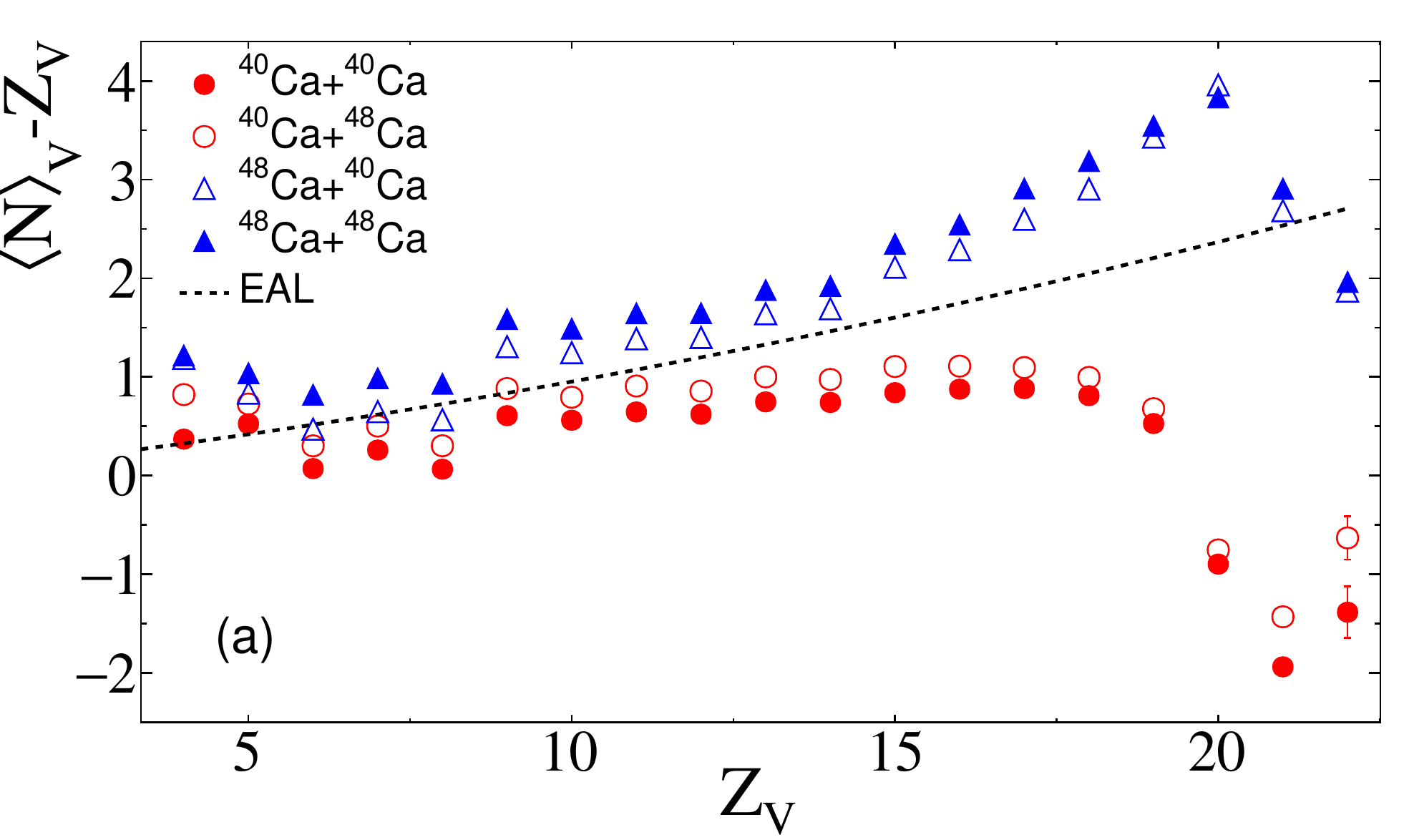}
\includegraphics[scale=0.42]{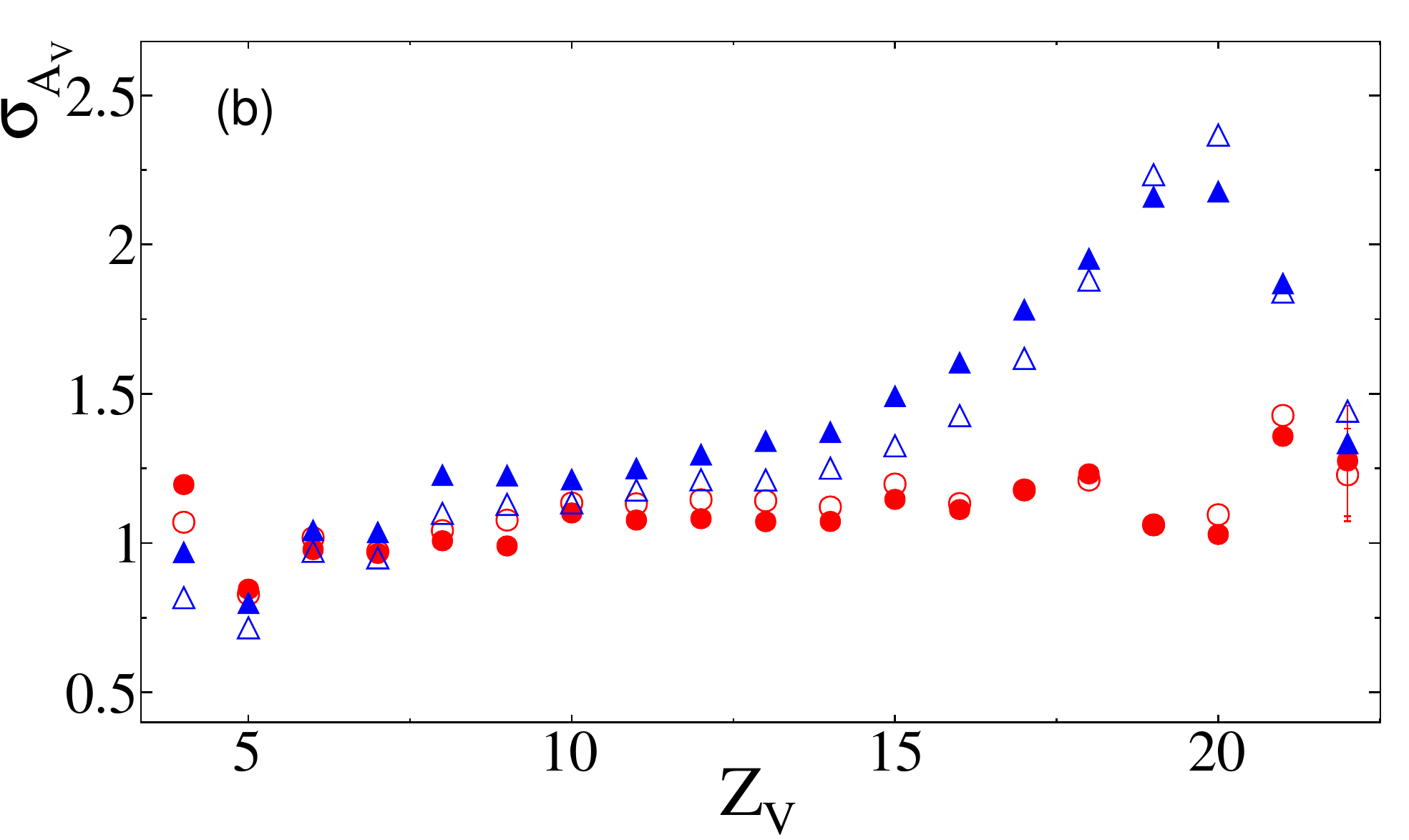}
\caption{(Color online) (a) Distribution of the average neutron excess of the fragments detected in VAMOS as a function of their atomic number for the four systems $^{40,48}$Ca+$^{40,48}$Ca at  35 $MeV$/nucleon. The dashed line represents the EAL parametrization (see text). 
(b) Associated isotopic distributions widths.}
\label{fig_plf_neutronexcess}
\end{figure}

In addition to the qualitative observations extracted from the charts of identified nuclides, Fig.\ref{fig_plf_neutronexcess}(a) shows the evolution of the average neutron excess of the fragments detected in VAMOS as a function of their atomic number for all the systems.
We observe an evolution of the curves according to the neutron content of the projectile: as expected, the fragments produced with $^{48}$Ca projectile systems are more neutron-rich compared to $^{40}$Ca projectile systems.
Furthermore, an effect of the target neutron richness is also visible (open symbols): for a given projectile, the neutron excess of all fragments is slightly higher for neutron-rich targets.
This can be interpreted as an experimental evidence of the isospin diffusion mechanism \cite{Shi2003:isospinDiffusionTheor,Tsang2004:isospindiffusion,Sun2010:isospinDiffusion,chbihi2018}.

For measured fragments close to the projectile, the $^{48}$Ca projectile reactions exhibit a mean neutron excess of $4$ compared to a mean neutron deficit of $1$ for the $^{40}$Ca projectile reactions (to be compared to the initial values $8$ and $0$ respectively). 
The average loss of four neutrons by the original $^{48}$Ca projectile nucleus may be explained by neutron transfer reactions between projectile and target, but also by the decay of an excited projectile preferentially via neutron emission. 
As a reminder, the residue corridor, or evaporation attractor line (EAL, here extracted from \cite{charity98} and represented in dashed line), is a region of the nuclear chart where proton and neutron emissions have equal probability at all excitation energies, acting as an attractor for decay chains.
Concerning the neutron-rich projectile systems, we observe that with decreasing charge of the fragment, the neutron excess decreases and the distributions merge to the EAL. This may reflect increasing excitation energy of the primary quasi-projectile fragments, which is also reflected in the reduced charge of the residual fragment.
Concerning the neutron-poor projectile systems, we observe that the mean neutron deficit of the fragment quickly decreases with its charge and the distributions also get closer to the EAL.
The aforementioned results appear as a direct measurement of the EAL, even though the evolution towards the line is governed by the interplay between isospin diffusion and secondary decays. Therefore it is difficult to disentangle both contributions.
Finally, it is worth noting that in the case of a pick-up process ($Z>Z_{proj}$) and for neutron rich systems, the addition of one or two charges is accompanied with a decrease of the neutron excess while for the neutron poor systems the neutron deficit remains approximately the same. 

Figure \ref{fig_plf_neutronexcess}(b) shows the evolution of the standard deviation of the isotopic distributions of the fragment identified in VAMOS. 
We can observe that the trends tend to follow the average neutron excess, with large values ($\sigma_{A}$ > 2) around $Z_V = 20$ for the $^{48}$Ca projectile systems.
This non-trivial experimental result is of interest as a relation between isotopic widths, isoscaling parameters and the symmetry energy can be established in various models, as long as the isotopic distributions can be approximated by Gaussians \cite{Ono_PRC_68_051601, Raduta2007}.
Nonetheless, a detailed study of the associated velocity spectra of such fragments with $Z=18-20$ reveals an overlap of contributions originating from several reaction mechanisms. 
This should be taken into account when using isotopic distributions of fragments close to the projectile to extract information on the symmetry energy.

\subsection{Characteristics of the light charged particles}

\subsubsection{Multiplicities}

In this subsection we are interested in the multiplicity of the light charged particles identified in INDRA in coincidence with the fragment in VAMOS. 
So as to focus on the particles most probably emitted by the decaying QP and reduce the contribution of pre-equilibrium and neck region emissions, a constant velocity cut of $V_{z}^{CM}>0$ in the center of mass frame is applied in this section. 

\begin{figure}
\includegraphics[scale=0.59]{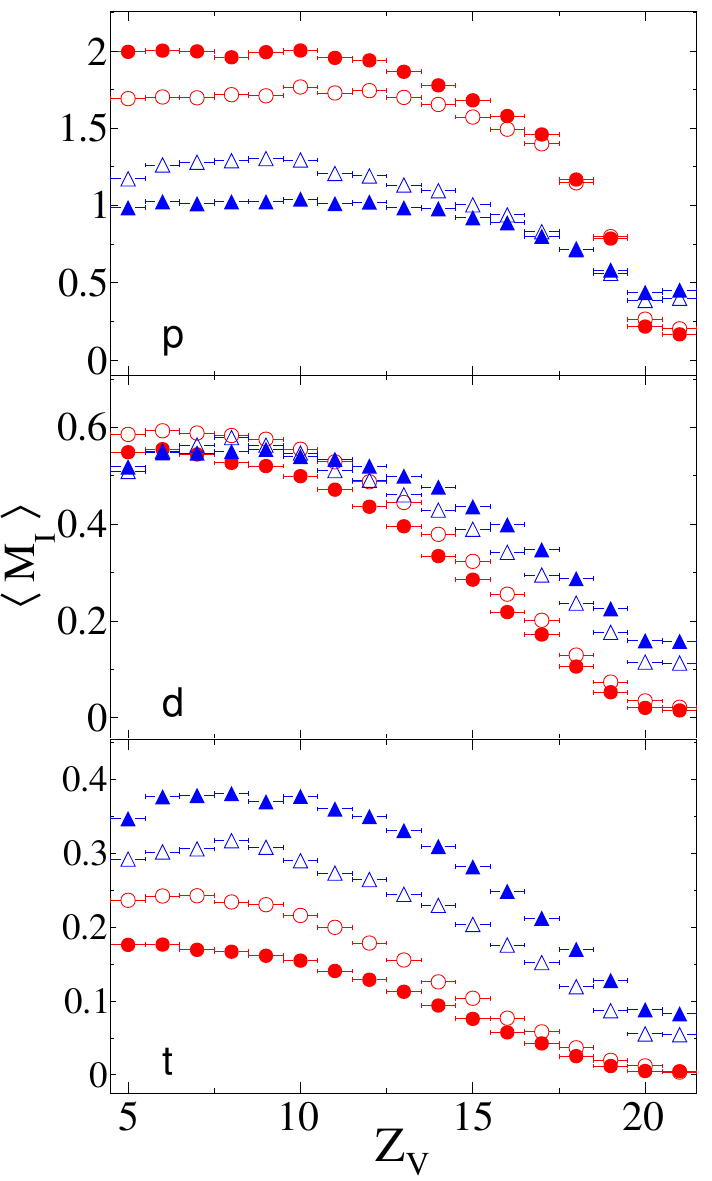}
\includegraphics[scale=0.59]{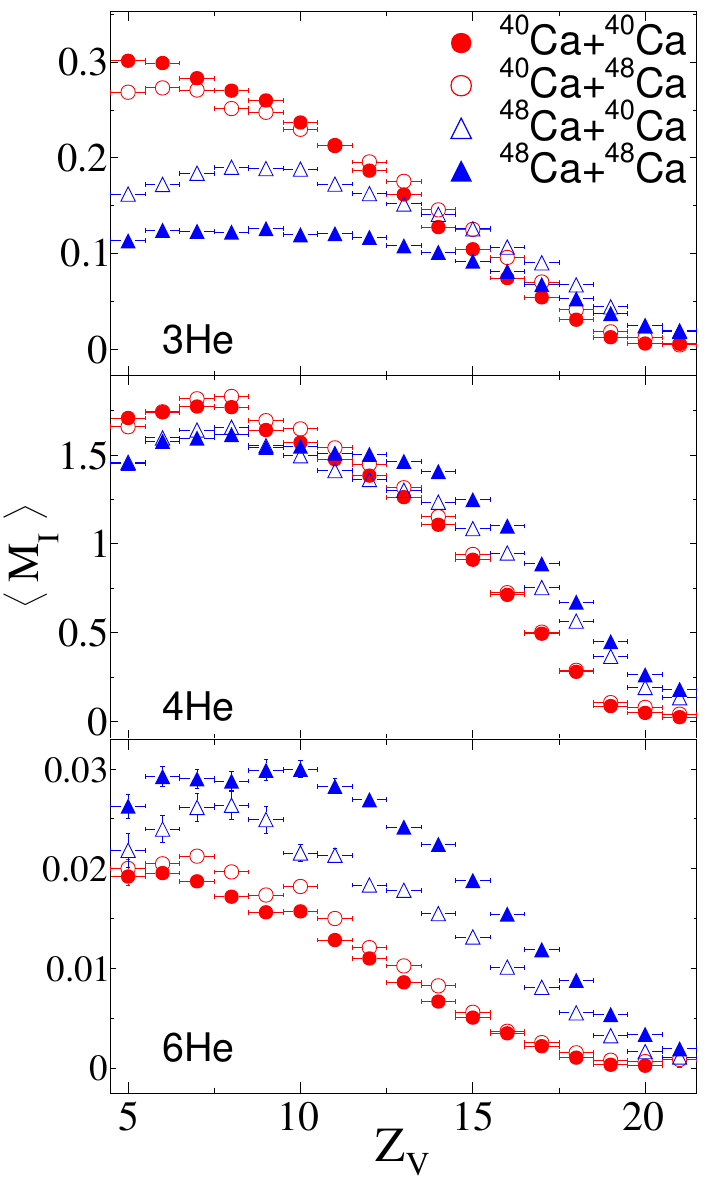}
\caption{Average forward-emitted ($V_{z}^{CM}>0$) LCP multiplicities detected in INDRA as a function of the charge $Z_{V}$ of the fragment identified in VAMOS, for the four $^{40,48}$Ca+$^{40,48}$Ca systems.}
\label{MultZplf}
\end{figure}

Figure \ref{MultZplf} shows the average multiplicity of the LCP identified in INDRA as a function of the atomic number of the fragment detected in VAMOS, $Z_{V}$.
First, we observe that the multiplicity increases with decreasing $Z_{V}$, reflecting increasingly dissipative collisions thus more excited systems. A saturation of the multiplicities is also observed for lower charges.
This is probably due to a much larger mixing of different impact parameters or degres of dissipation for such small $Z_V$ values. 
Second, by comparing the four systems we observe a trend according to the neutron enrichment. Protons and neutron-poor $^{3}He$ particles, but also neutron-rich tritons and $^{6}He$ particles, show multiplicities that are related first to the neutron richness of the projectile then, to a lesser extent, to the one of the target as the centrality increases.
Third, the multiplicities of deuterons and $^{4}He$, particles having the same number of protons and neutrons, present a different trend compared to the previous observations and their multiplicities depend much less on the system. 
The $^{48}$Ca projectile systems exhibit $^{4}He$ and deuteron multiplicities higher than the $^{40}$Ca projectile case for $Z_{V} \geq 13$. 
The opposite trend is observed for $Z_{V}<13$, where the $^{4}He$ and deuteron emission is enhanced for neutron-deficient systems.
Again, it is highly unlikely that the same values of $Z_V$ for the different systems correspond directly to the same impact parameters or degrees of dissipation.
The aformentioned observations are in agreement with previously published results of the INDRA collaboration, focusing on the study of the LCP emitted in $^{136,124}$Xe+$^{124,112}$Sn collisions \cite{BOUGAULT_PhysRevC_97_024612}.

Finally, the low values of the average LCP multiplicities observed in Fig.\ref{MultZplf} indicate that the parent of the associated PLF is moderately excited.
The minima observed around $Z_{V}=Z_{proj}$ could indicate that the excitation energy of the primary fragment is minimal in this region. 

\subsubsection{Kinematical properties of the LCP}

Some kinematical properties of the LCP can be observed from the invariant cross section contours in the parallel versus longitudinal velocity plots ($V_{\parallel}-V_{\perp}$).

\begin{figure}
\includegraphics[scale=0.38]{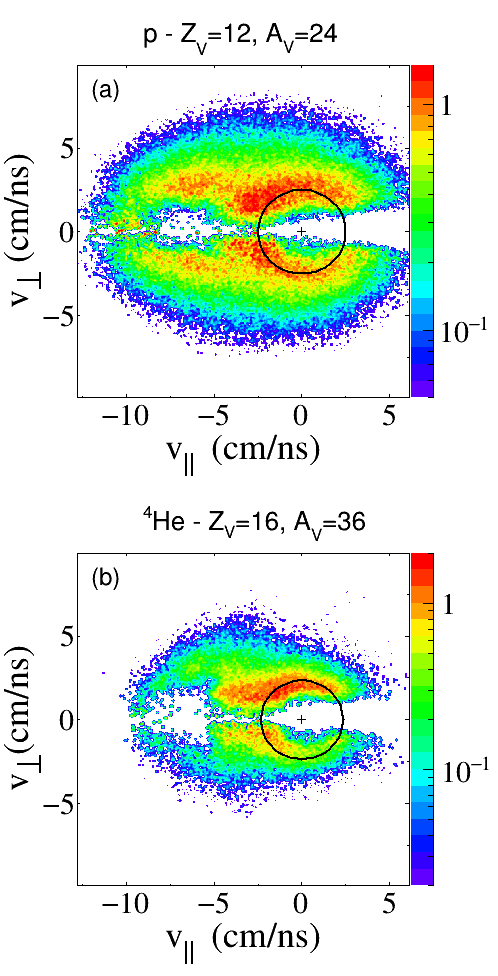}
\caption{(Color online) 
Invariant $V_{\parallel}-V_{\perp}$ maps in the VAMOS fragment frame (a) for protons emitted in coincidence with a $^{24}$Mg for the $^{40}$Ca$+^{40}$Ca system and (b) alpha particles emitted in coincidence with a $^{36}$S for the $^{40}$Ca$+^{40}$Ca system.
The circles represent the most probable velocities of the particles evaporated from a $^{48}$Ca nucleus \cite{PhysRevC_44_774}.} 
\label{fig_vpar_vper_costheta}
\end{figure}

Fig.\ref{fig_vpar_vper_costheta}(a) and (b) show such plots in the VAMOS fragment frame, respectively for protons and $^{4}He$ particles emitted in coincidence with two different isotopes with the $^{40}$Ca$+^{40}$Ca and $^{48}$Ca$+^{48}$Ca systems. 
We clearly observe two main components, one centered at $V_{\parallel} = 0$ and the other at $V_{\parallel} \simeq -8$ $cm/ns$, respectively corresponding to the QP and QT sources.
The presence of an overlap at mid-rapidity ($V_{\parallel} \simeq -4$ $cm/ns$, corresponding to the center of mass parallel velocity) is also visible.

\section{Experimental study of the isoscaling method}\label{sec_csym}

\subsection{The isoscaling method}\label{subsec_isoscaling}

Information on the symmetry energy coefficient of the binding energy of finite nuclei $C_{sym}$ can be inferred from the scaling behaviour, also called isoscaling, obtained from the ratio $R_{21}(N,Z)$ of the yields of the same isotope measured with two systems, $Y_{(1)}(N,Z)$ and $Y_{(2)}(N,Z)$, where (2) usually stands for the neutron-rich system \cite{Tsang2001:symmetryEnergy}. 
Indeed, in a variety of HIC an exponential dependence of the ratio on $N$ and $Z$ has been observed, such as:
\begin{equation}
\label{eq_iso_params}
R_{21}(N,Z) = \frac{Y_{(2)}(N,Z)}{Y_{(1)}(N,Z)} \propto exp\left[ \alpha N + \beta Z \right] 
\end{equation}
where $\alpha$ and $\beta$ are called the isoscaling parameters.

Assuming that a set of carefully selected fragmenting sources can be described in the approximation of a grand canonical statistical ensemble, $\alpha$ and $\beta$ can be expressed as $\alpha=\Delta \mu_{n}/T$ and $\beta=\Delta \mu_{p}/T$, where $\Delta \mu_{n}$ and $\Delta \mu_{p}$ are the differences between the neutron and proton chemical potentials and $T$ the temperature of the decaying systems \cite{Ono_PRC_68_051601}.

Furthermore, a Gaussian approximation of the yields in the grand-canonical approximation allows to link the $\alpha$ parameters to the symmetry energy coefficient divided by the temperature for a given fragment charge $Z$, such as \cite{Friedman1988PhysRevLett_60_2125, Friedman1990PhysRevC_42_667, Tsang2001:symmetryEnergy, BotvinaPhysRevC_65_044610, Ono_PRC_68_051601, Raduta2007, PhysRevC_75_024605}:
\begin{equation}
\label{eq_csymalpha}
\frac{4C_{sym}(Z)}{T}=\frac{\alpha(Z)}{(\frac{Z}{\langle A_{1}(Z) \rangle})^{2}-(\frac{Z}{\langle A_{2}(Z) \rangle})^{2}}
\end{equation}
where $\langle A_{1} \rangle$ and $\langle A_{2} \rangle$ are the mean masses corresponding to the isotope charge $Z$ for each reaction.

In this context, the following sections present the suitability of the isoscaling method applied to the experimental data measured with INDRA-VAMOS, along with a reconstruction method used to estimate the characteristics of the quasi-projectile from the measured reaction products.  

\subsection{Primary fragment reconstruction method}\label{subsec_QP_recon}

Several studies of the isotopic properties of the primary and secondary fragment yield distributions lead to the conclusion that the latter and the associated neutron-to-proton ratio are affected by the secondary deexcitation effects \cite{PhysRevC_71_024602, PhysRevC_79_061602, PhysRevC_98_044602}.
More specifically, the quality of the isoscaling fits, but also the average fragment isotopic composition, are expected to be distorted by secondary decays.

We have applied a reconstruction method to explore these effects and estimate the warm primary QP thermodynamical properties, by appropriately selecting the isotopically identified particles and fragments detected in INDRA, in coincidence with the fragments measured in VAMOS \cite{PhysRevC_98_044602}.

In order to isolate the QP emissions, we have used the relative velocities between the reaction products detected in INDRA and (i) the PLF detected in VAMOS ($V_{rel,PLF}$), (ii) the fragment with the largest identified Z at backward angles in INDRA, supposed to be the TLF ($V_{rel,TLF}$). 
Numerically, cuts on the associated relative velocities, respectively $V_{rel,PLF}$ and $V_{rel,TLF}$, were applied so as to include fragments whose relative velocities verify $\frac{V_{rel,TLF}}{V_{rel,PLF}} > 1.35$ for $Z=1$ and $\frac{V_{rel,TLF}}{V_{rel,PLF}} > 1.75$ for $Z \geq 2$.
In case no fragment was identified in INDRA at backward angles, the overall mean velocity of the TLF for a given $Z_{V}$ was used to compute $V_{rel,TLF}$.
The values of the cut-off threshold were first estimated from the AMD calculations, by comparing the parallel velocities $V_z^{V}$ distributions of the accepted nuclei in the VAMOS fragment frame, to the one effectively evaporated from the QP in AMD. 
As a second step, the aformentioned cuts were optimized in order to reproduce as best as possible the excitation energy per nucleon, charge, mass and isotopic distributions of the QP within AMD.
Within the filtered model, the actual quasi-projectile contribution to the selected $Z=1$ nuclei ranges from $60\%$ to $70\%$ for $Z_{V}=5-20$.
Such values are comparable with the one obtained from HIPSE-SIMON and CoMD for the $^{64}$Zn$+^{64}$Zn, $^{70}$Zn$+^{70}$Zn and $^{64}$Ni$+^{64}$Ni reactions with the NIMROD detector \cite{MARINI201380}.

\begin{figure}
\includegraphics[scale=0.45]{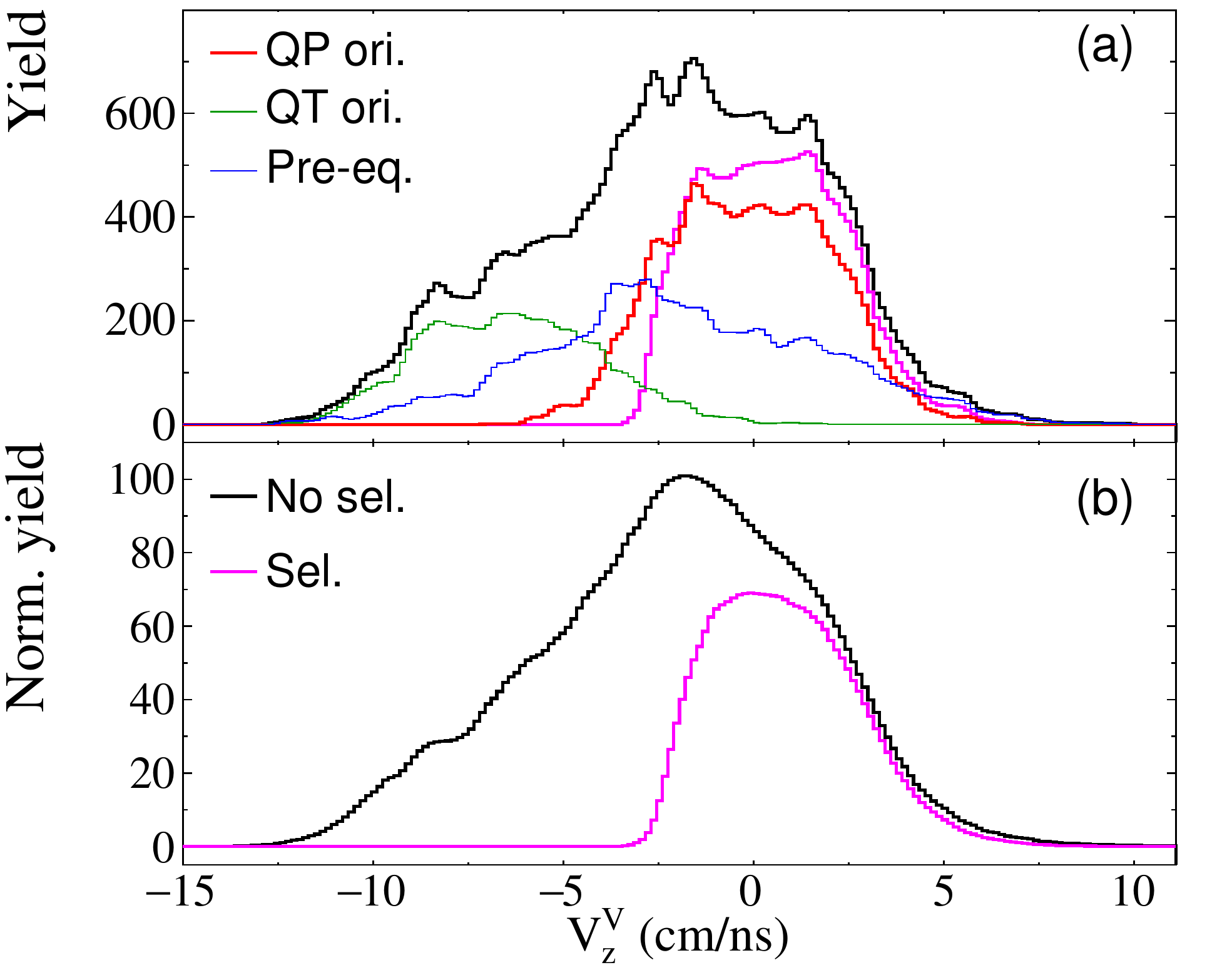}
\caption{(Color online) Selection of the QP source for the $Z=1$ isotopes with $Z_{V}=15$ for the $^{40}$Ca+$^{48}$Ca system (see text). 
(a) Effect of the selection for the filtered model calculations.
(b) Effect of the selection for the data.}
\label{fig_vpar_sel}
\end{figure}

It is worth noting that the effect of the cuts are consistent between the experiment and the model. 
An example is given Fig.\ref{fig_vpar_sel} for the $Z=1$ isotopes, where we observe a non-negligible contribution of the QT (green) and pre-equilibrium (blue) on the overall velocity distribution (black), while the selection (magenta) tends to reproduce the QP source distribution (red).
Moreover, an earlier work in the framework of the semi-classical Landau-Vlasov approach has shown similar dynamical effect in HIC at intermediate energies over the whole impact parameter range \cite{PhysRevC_56_2003}. 

The proposed method turned out to be necessary to isolate the QP emissions. 
Indeed, according to the filtered model calculations and on the contrary to the results of \cite{PhysRevC_98_044602}, cuts on $V_{rel,PLF}$ only are not sufficient to isolate the QP emissions, neither is a constant velocity cut of $V^{CM}_{z}>0$ in the center of mass frame. 
For the reactions under study, such selections would lead to an overestimation of about 1 MeV/nucleon of the average excitation energy for the reconstructed QP (presented in Section \ref{subsec_calo}).

\begin{figure}
\includegraphics[scale=0.44]{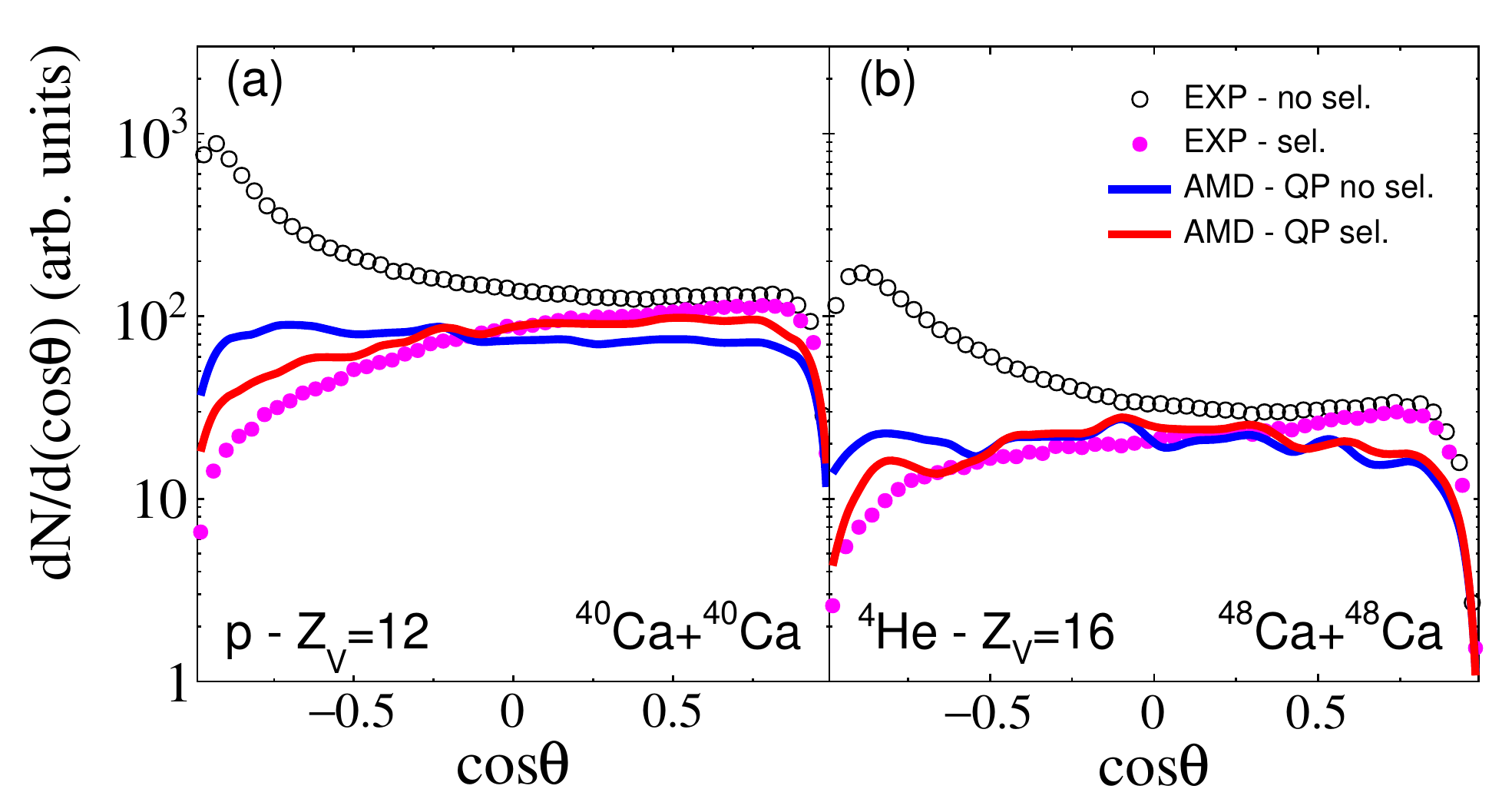}
\caption{(Color online) $cos(\theta)$ distributions in the VAMOS fragment frame from the experiment with (full symbols) or without (open symbols) the velocity selection and from the particles only emitted by the QP from the filtered model calculations (continuous lines). 
(a) Distribution for the protons with $Z_{V}=12$ for the $^{40}$Ca$+^{40}$Ca reaction.
(b) Distribution for the $^{4}He$ particles with $Z_{V}=16$ for the $^{48}$Ca$+^{48}$Ca reaction.
The model distributions were normalized to the integral of the corresponding selected experimental distributions for the plot.}
\label{fig_costheta_sel}
\end{figure}

Fig.\ref{fig_costheta_sel} shows an example of the resulting $cos(\theta)$ distributions for protons and $^{4}He$ particles, before and after application of the velocity selection on the data.
A comparison with AMD is also given, for filtered particles only emitted by the QP whose remnant passes the VAMOS filter. 
We observe that the effect of the cut is, as expected, to reduce the contribution of LCP emitted at backward angles, thus emitted by other sources than the QP. 
We can notice a good agreement of the selection between the experiment and the model, meaning that the velocity cut is suitable to isolate protons and alpha particles emitted by a decaying QP. Nonetheless, the selection tends to introduce an anisotropy in the model distributions in the $cos(\theta)< -0.5$ region.

The quasi-projectile atomic number $Z_{QP}$ is reconstructed as the sum of the atomic numbers of the fragment detected in VAMOS and the $M_{I}$ identified and selected LCP and IMF detected in INDRA in the event, such as:

\begin{equation}\label{eq_ZQP}
Z_{QP} = Z_{V} + \sum_{i}^{M_{I}} Z_{i}
\end{equation}
where $Z_{V}$ and $Z_{i}$ are respectively the charges of the fragment measured in VAMOS and accepted evaporated nucleus $i$. 

The associated quasi-projectile mass number without the evaporated neutron contribution, $\widetilde{A}_{QP}$, is reconstructed as the sum of the mass numbers of the fragment in VAMOS and the $M_{I}$ identified and accepted LCP and IMF detected in INDRA, such as:
\begin{equation}\label{eq_AQPwon}
\widetilde{A}_{QP}=A_{V}+\sum_{i}^{M_{I}} A_{i} 
\end{equation}

As stated in \cite{PhysRevC_79_061602, PhysRevC_98_044602}, the quality of the isoscaling fits could be greatly improved by including the neutrons evaporated by the QP. 
As the neutrons were not measured for the present experiment, the distributions of the neutrons evaporated by the reconstructed QP from the filtered model calculations were used as a substitute.
More precisely, for each event with reconstructed charge $Z_{QP}$ and mass without the neutron $\tilde{A}_{QP}$, the experimental evaporated neutron multiplicity was estimated from a random number generator following the filtered model neutron multiplicity distribution (histogram).
A scaling factor was also applied in order to take into account the fact that the model systematically overestimates the light particle multiplicities. This is due to the AMD version used for the analysis, which does not include cluster correlations introduced in a more recent version of the model \cite{Ono_2013}.
Assuming that the experimental and filtered model average neutron-to-proton multiplicity ratios are equivalent for each $Z_{QP}$, we have:
\begin{equation}\label{eq_kratio}
\langle M_n \rangle^{exp} = \langle M_n \rangle^{mod} \cdot \frac{\langle M_p \rangle ^{exp}}{\langle M_p \rangle ^{mod}} = \langle M_n \rangle^{mod} \cdot k 
\end{equation}
where $\langle M_{n,p} \rangle^{exp,mod}$ are the neutron and proton average multiplicities of the experiment and the model. 
The experimental evaporated neutron multiplicity is:
\begin{equation}
\label{eq_mn}
M_n (Z_{QP}, \tilde{A}_{QP}) = \lceil M_n^{rdm}(Z_{QP}, \tilde{A}_{QP}) \cdot k \rceil
\end{equation}
where $M_n^{rdm}$ is the random neutron multiplicity extracted from the model histogram and $\lceil \rceil$ is the ceiling function.
A constant value of $k=\langle k \rangle = 0.7$, corresponding to the average scaling factor over all systems and charges, was applied to the data.
It is important to note that this work is not focused on the fine tuning of AMD parameters, explaining why a fixed value is used for $k$.

The experimental reconstructed QP mass $A_{QP}$ is thus:

\begin{equation}\label{eq_AQP}
A_{QP} = \widetilde{A}_{QP} + M_n (Z_{QP}, \widetilde{A}_{QP})
\end{equation}

\begin{figure}
\centering
\includegraphics[scale=0.44]{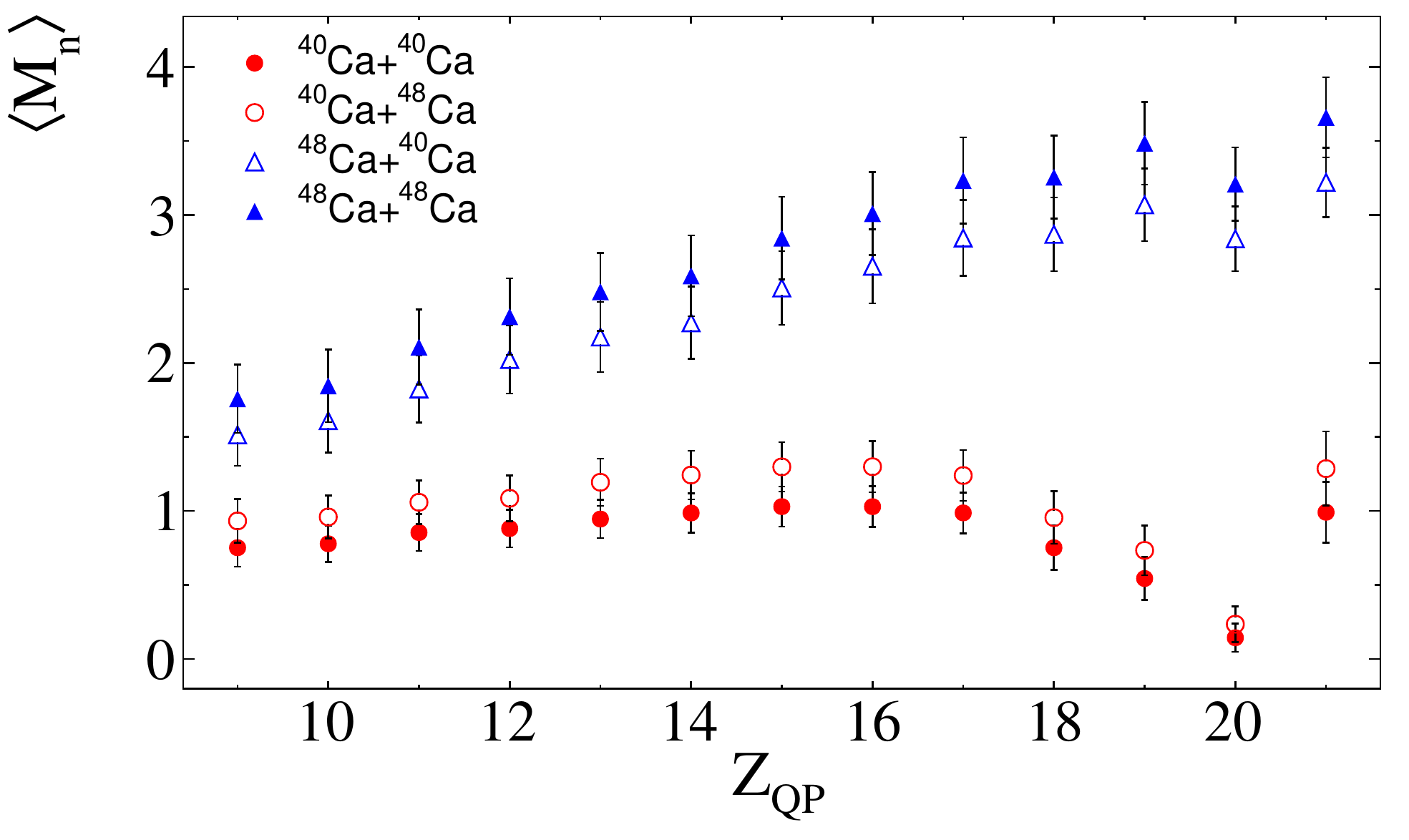}
\caption{(Color online) Average neutron multiplicities distribtions used for the mass reconstruction of the quasi-projectile as a function of its charge.}
\label{fig_nmult}
\end{figure}

The corresponding experimental average neutron multiplicity distributions as a function of $Z_{QP}$ are presented in Figure \ref{fig_nmult} for the four systems under study.

\begin{figure}
\centering
\includegraphics[scale=0.44]{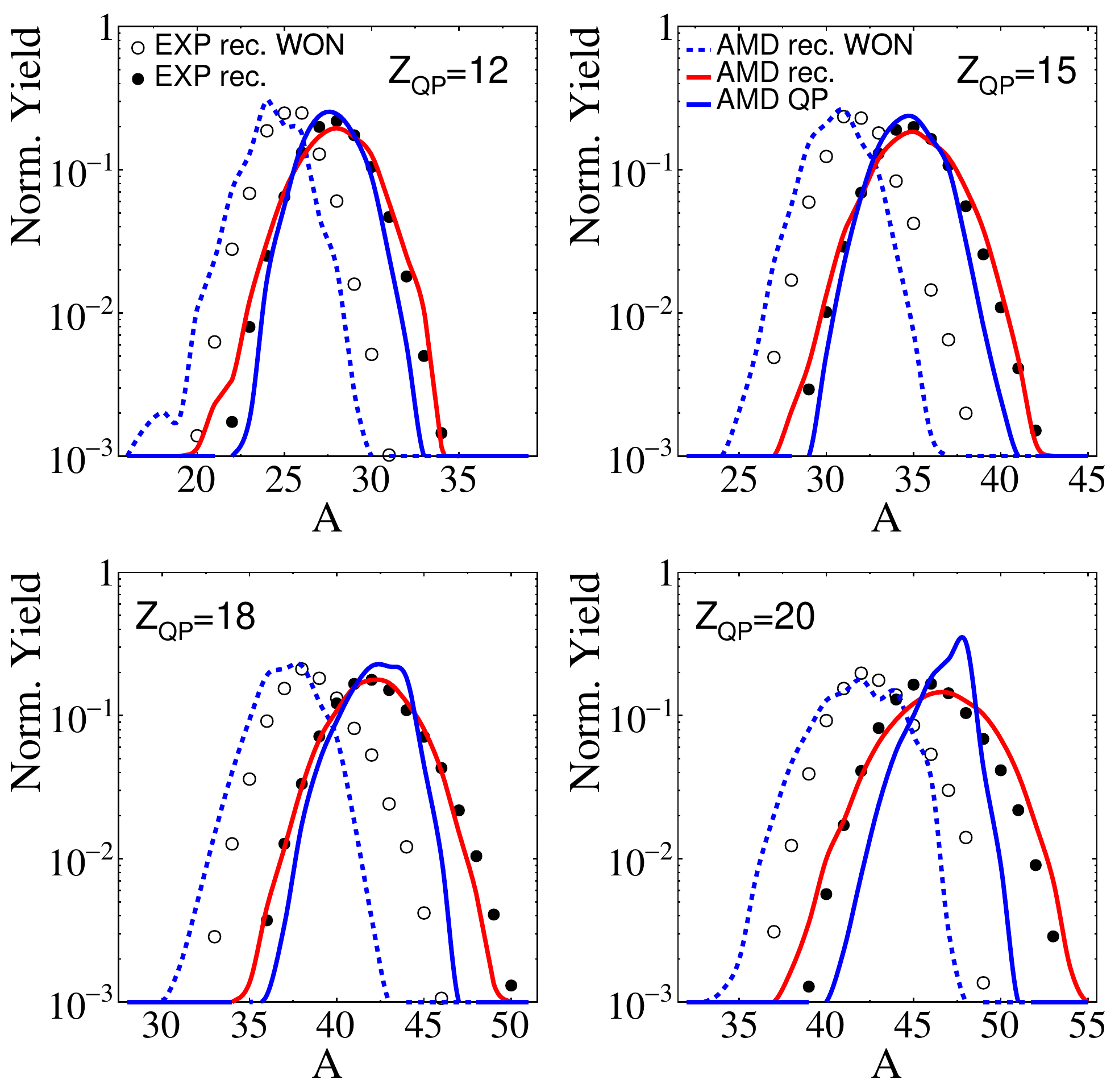}
\caption{(Color online) Comparison of QP isotopic distributions for the experiment and the filtered model calculations, for the $^{48}$Ca$+^{48}$Ca system.
The reconstructed QP without the evaporated neutron contribution (WON) are represented for the model and the experiment by dashed lines and open circles  resepectively, while the results with the neutron contribution are represented by solid blue lines and full circles.
The actual QP from AMD is represented by continuous red lines.}
\label{fig_AQPtilde_AQP_recon_exp_4848}
\end{figure}

An example of the reconstructed isotopic distributions is given in Fig.\ref{fig_AQPtilde_AQP_recon_exp_4848} and compared to the model for some atomic numbers of the QP for the neutron-rich $^{48}$Ca$+^{48}$Ca reaction.
Within the model, the neutron contribution seems mandatory to better reproduce the actual QP distribution with the reconstruction.
Concerning the reconstructed distributions without the neutrons, we observe that the model (dashed lines) reproduces quite well the corresponding experimental mean values (open circles). 
The actual QP isotopic distributions from the filtered model are also represented (solid lines). We observe that the neutron corrected experimental data tend to reproduce the mean values for $Z_{QP} \neq 20$ but exhibit higher widths than the QP from the model.
Comparing the actual QP distributions with the reconstructed one corrected from the neutrons within the model in the $Z_{QP} = 8-21$ range, a mass difference decreasing from $-0.5$ to $0.5$ with increasing $Z_{QP}$ is observed, while it varies from $-0.5$ to $0$ for the projectile $^{40}$Ca reactions.
In the same range, the relative deviations of the widths increase from $5\%$ to $20\%$ for the neutron-rich projectile $^{48}$Ca reactions and from $5\%$ to $10\%$ for the projectile $^{40}$Ca reactions.

It should also be noted that several other methods based on a random selection of the neutron multiplicities from their associated distributions (filtered, with different $Z_{QP}$, $Z_{V}$ or/and $\widetilde{A}_{QP}$, $A_{V}$ combinations) were tested. 
These methods were discarded as they often introduce spurious discrepancies in the resulting isotopic distributions, with larger mean and width deviations.

\subsection{Evolution of the isoscaling parameters}\label{subsec_alpha_study}

Starting from this section, the fragment identified in VAMOS is assumed to be the PLF.

It must be noted that an additional offline condition is applied in order to remove the events with a PLF measured in INDRA, such as:
\begin{equation}\label{eq_cut_Zv_ZHfwd}
Z_{V} > Z_{I}^{max,fwd}
\end{equation}
where $Z_{I}^{max,fwd}$ is the charge of the forward-emitted ($V_{z}^{CM}>0$) heaviest fragment identified in charge with INDRA.
Indeed, such events are not relevant for the study of isoscaling and isospin-sensitive observables as the mass-identification in INDRA is limited to $Z\simeq5$.
Less than $2\%$ of the whole statistics is removed by this selection for all the systems under study. 

The experimental isoscaling parameters $\alpha$ were extracted from linear fits applied to the natural logarithm of the yield ratios $R_{21}$ of Eq.\ref{eq_iso_params} for each charge $Z$, across all PLF and reconstructed QP isotopes. More details about the fitting procedure are given in Appendix \ref{app3_isofits}.

\begin{figure}[ht]
\centering
\includegraphics[scale=0.52]{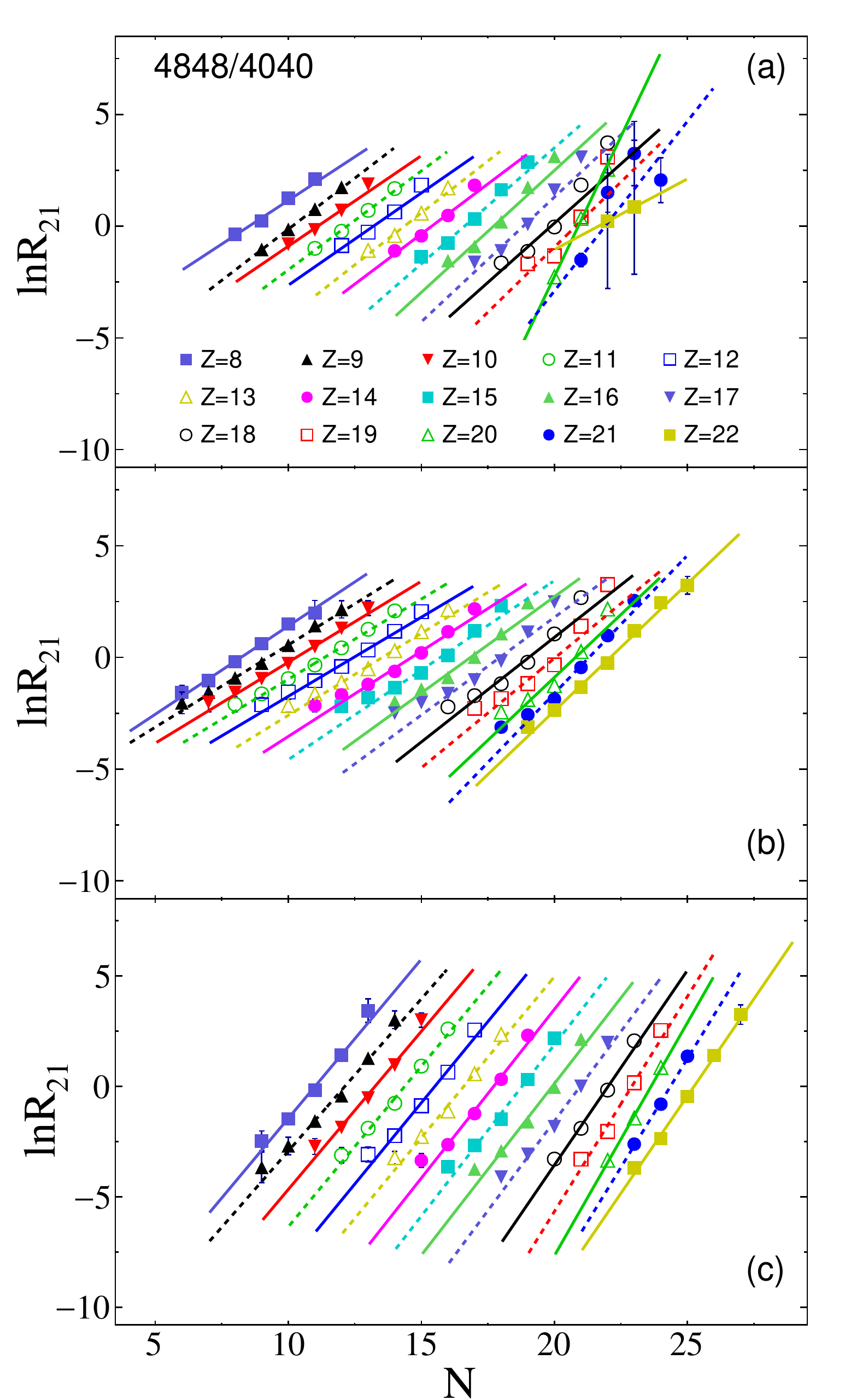}
\caption{(Color online) Example of experimental isoscaling plots using an expanded $Z$ range for the $^{48}$Ca$+^{48}$Ca system relative to the $^{40}$Ca$+^{40}$Ca with the PLF (a), the reconstructed QP without (b) and with (c) the evaporated neutron contributions.
The solid and dashed lines correspond to the resulting fits according to Eq.\ref{eq_iso_params} for even and odd charges respectively.}
\label{fig_fit_alpha_iso_recon}
\end{figure}

An example of the experimental fits for the PLF, the reconstructed QP without or with the evaporated neutrons contribution is given in Fig.\ref{fig_fit_alpha_iso_recon}, for the $^{48}$Ca+$^{48}$Ca system relative to the $^{40}$Ca+$^{40}$Ca. 
Each symbol represents a given isotope while the fits of Eq.\ref{eq_iso_params} to the ratios are represented with solid and dashed lines.
We can observe, for all cases, that the yield ratios exhibit a clear isoscaling in the $Z=8-16$ range, while it tends to disappear for higher charges for the PLF.
The isoscaling is verified for higher charges in the case of the reconstructed QP, with a noteworthy change in the slope if the evaporated neutron contribution is considered. 
Nonetheless, we observe a discontinuity of the fits for $Z=19-20$, explained by the overlap of contributions originating from several reaction mechanisms, as discussed in Sec. \ref{subsec_QP_recon}.
%
%

\begin{figure}[ht]
\centering
\includegraphics[scale=0.57]{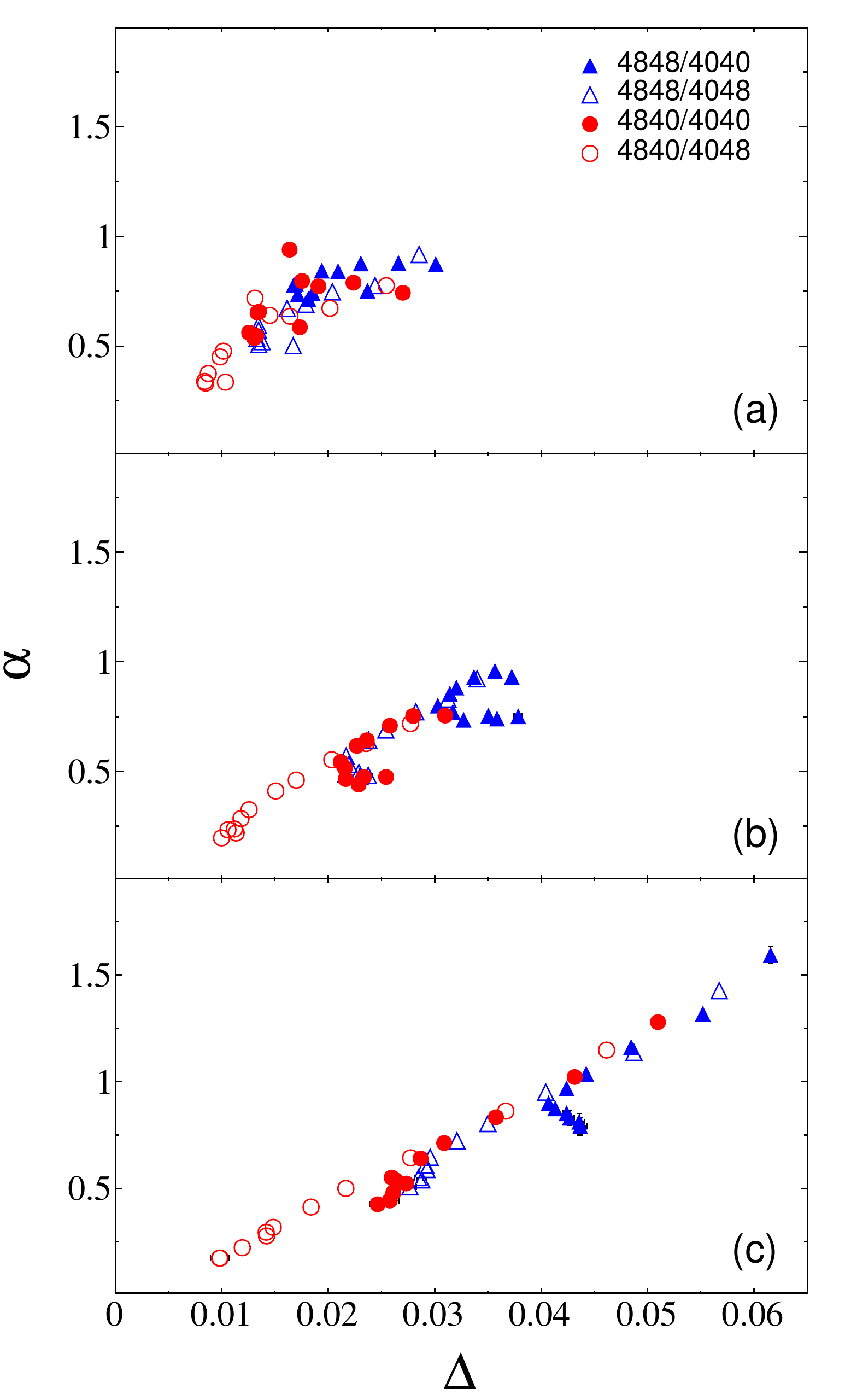}
\caption{(Color online) Experimental isoscaling $\alpha$ parameters as a function of the difference in average neutron composition of the two sources in the $Z=9-19$ range, for the four combinations under study with the PLF (a), the reconstructed QP without (b) and with (c) the evaporated neutron contributions. The values are averaged over $k=0.6-0.8$ (see text) and the propagated errors are smaller than the size of the points.}
\label{fig_alpha_vs_delta}
\end{figure}

For more consistency, we have also studied the sensitivity of the isoscaling parameters to the evaporated neutrons estimation, by varying the scaling factor $k$ (see Eq.\ref{eq_kratio}) from $k=0.6-0.8$. 
This domain corresponds to a variation of a one standard deviation of the neutron multiplicities obtained for $\langle k \rangle = 0.7$. 
The results presented thereafter are averaged over this domain in $k$.
The overall dependence of $\alpha$ on the difference in average neutron composition of the two sources $\Delta = (Z / \langle A_{1} \rangle)^{2}-(Z / \langle A_{2} \rangle)^{2}$ (see Eq. \ref{eq_csymalpha}) is shown in Fig.\ref{fig_alpha_vs_delta}, for the $Z=9-19$ range where the isoscaling is verified.

According to the literature, the magnitude of the $\alpha$ parameter is expected to linearly increase with increasing difference in the asymmetry of the two sources $\Delta$ \cite{PhysRevC_70_011601, PhysRevC_79_061602}. 
Such correlation is not observed with the present data in the case of the PLF, while it is visible for the reconstructed QP. 
Comparing Fig.\ref{fig_alpha_vs_delta}(a) and Fig.\ref{fig_alpha_vs_delta}(b), we can conclude that the secondary de-excitation effect tends to not only lower the experimentally observed values but also remove the correlation between the two parameters, making the isoscaling of the PLF not relevant for the reactions under study.
As seen in Fig.\ref{fig_alpha_vs_delta}(c), the correction for evaporated neutrons leads to an increase of the values while the linearity is still observed.
Small deviations are nonetheless observed, more pronounced with the neutron-rich $^{48}$Ca+$^{48}$Ca system. 
These correspond to the low QP charge region ($Z_{QP}<12$) where we expect the reconstruction method to be less relevant as it is harder to determine that the fragment detected in VAMOS is indeed a PLF.  

\begin{figure}[ht]
\centering
\includegraphics[scale=0.57]{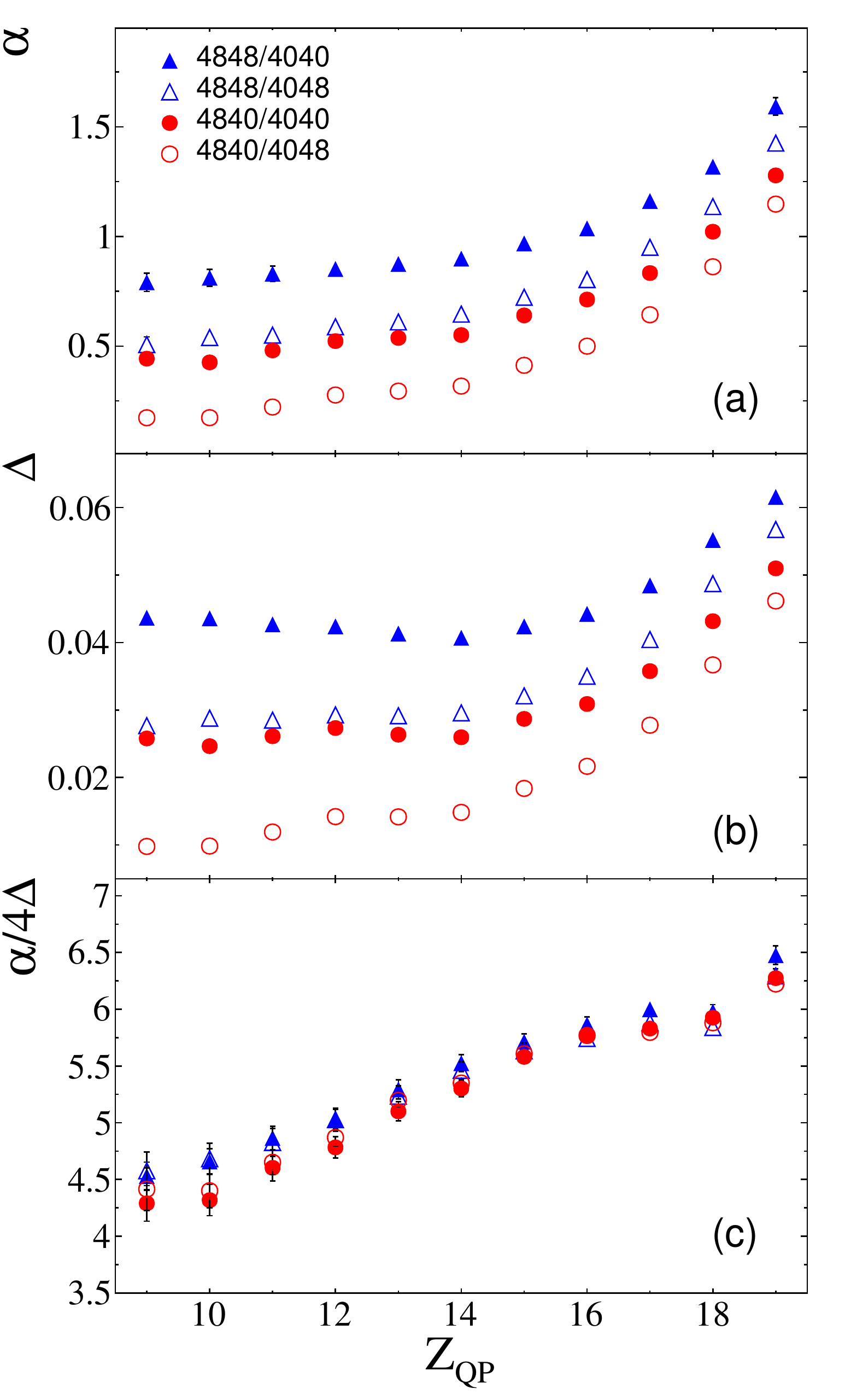}
\caption{(Color online) 
(a) Experimental isoscaling $\alpha$ parameter as a function of the reconstructed charge with the evaporated neutron contribution.
(b) Associated difference in average neutron composition of the two sources.
(c) Corresponding ratio $\alpha/4\Delta = C_{sym}/T $.
The values are averaged over $k=0.6-0.8$.}
\label{fig_alpha_denum_csymt}
\end{figure}

Fig.\ref{fig_alpha_denum_csymt}(a) and \ref{fig_alpha_denum_csymt}(b) present respectively the experimental $\alpha$ and $\Delta$ parameters as a function of the charge of the reconstructed QP with the evaporated neutron correction. 
A clear increase of both parameters is observed with the size of the reconstructed QP ($Z_{QP}$), which could be interpreted as an experimental evidence of a strong surface dependence of the symmetry energy term \cite{PhysRevC_75_024605}.
At first sight, this behaviour seems in opposition with the results reported in \cite{souliotis2003:isotopicScaling, SOULIOTIS200435, PhysRevC_75_011601, PhysRevC_90_064612}, where decreasing $\alpha$ values are observed with increasing charge.
It is in fact possible to reproduce such behaviour with the present data by using combinations of systems with the same projectile in Eq.\ref{eq_iso_params}, similarly to the studies of Souliotis \textit{et.~al.}.
The associated error bars are nonetheless significantly larger (by a factor of 3) than the ones obtained with the four combinations presented in Fig.\ref{fig_alpha_denum_csymt}.
It is also worth noting that the overall values of $\alpha$ and $\Delta$ are compatible with the one obtained from $^{86,78}$Kr$+^{64,58}$Ni reactions at $35$ MeV/nucleon, measured with the NIMROD-ISiS array \cite{PhysRevC_79_061602}.
Nonetheless, Wuenschel \textit{et.~al.} analysis focuses on $Z=1-17$ isotopes (for complete events) while the projectile charge is $Z=36$, we thus expect the impact parameter domain to be rather different.
This could explain why the $\alpha$ parameters obtained in \cite{PhysRevC_79_061602} are not charge dependent.

Finally, we observe for both parameters a similar hierarchy according to the system combination used with the isoscaling method.
This hierarchy is an experimental evidence that justifies the use of the isoscaling $\alpha$ parameter as a surrogate for isospin asymmetry in isospin transport studies \cite{PhysRevC_98_044602}.
%

To go further, using Eq.\ref{eq_csymalpha} we can study the behaviour of the symmetry energy term from the ratio of the extracted $\alpha$ and $\Delta$ parameters.
The extracted $\alpha/4\Delta = C_{sym}/T$ values are presented as a function of the charge of the reconstructed QP in Fig.\ref{fig_alpha_denum_csymt}(c).
Similarly to Fig.\ref{fig_alpha_denum_csymt}(a) and (b), an increase of $\alpha/4\Delta$ with the charge, thus the size of the QP, is observed for all available combinations with relatively close values for all the considered combinations.
The change in temperature with $Z_{QP}$ must nonetheless be understood to draw conclusion about the symmetry energy term itself.

\subsection{Excitation energy and temperature estimation}\label{subsec_calo}

The reconstruction of the QP also allows to estimate its excitation energy using calorimetry, such as:

\begin{equation}\label{eq_calo}
E^{*} = \sum_i^{M_{CP}} Ek_i + M_{n} \cdot \langle Ek_n \rangle - Q
\end{equation}
where $E^{*}$ is the excitation energy of the QP, $Ek_i$ the kinetic energy of the identified and selected charged particle $i$ in the rest frame of the reconstructed QP, $M_n$ the estimated neutron multiplicity, $\langle Ek_n \rangle$ the average neutron kinetic energy and $Q$ the mass balance of the reconstruction of the QP from the PLF, the accepted charged particles and the estimated neutrons. 
The neutrons average kinetic energy was computed using the proton one with a correction for the Coulomb barrier energy \cite{Vaz1984}.

\begin{figure}[ht]
\centering
\includegraphics[scale=0.57]{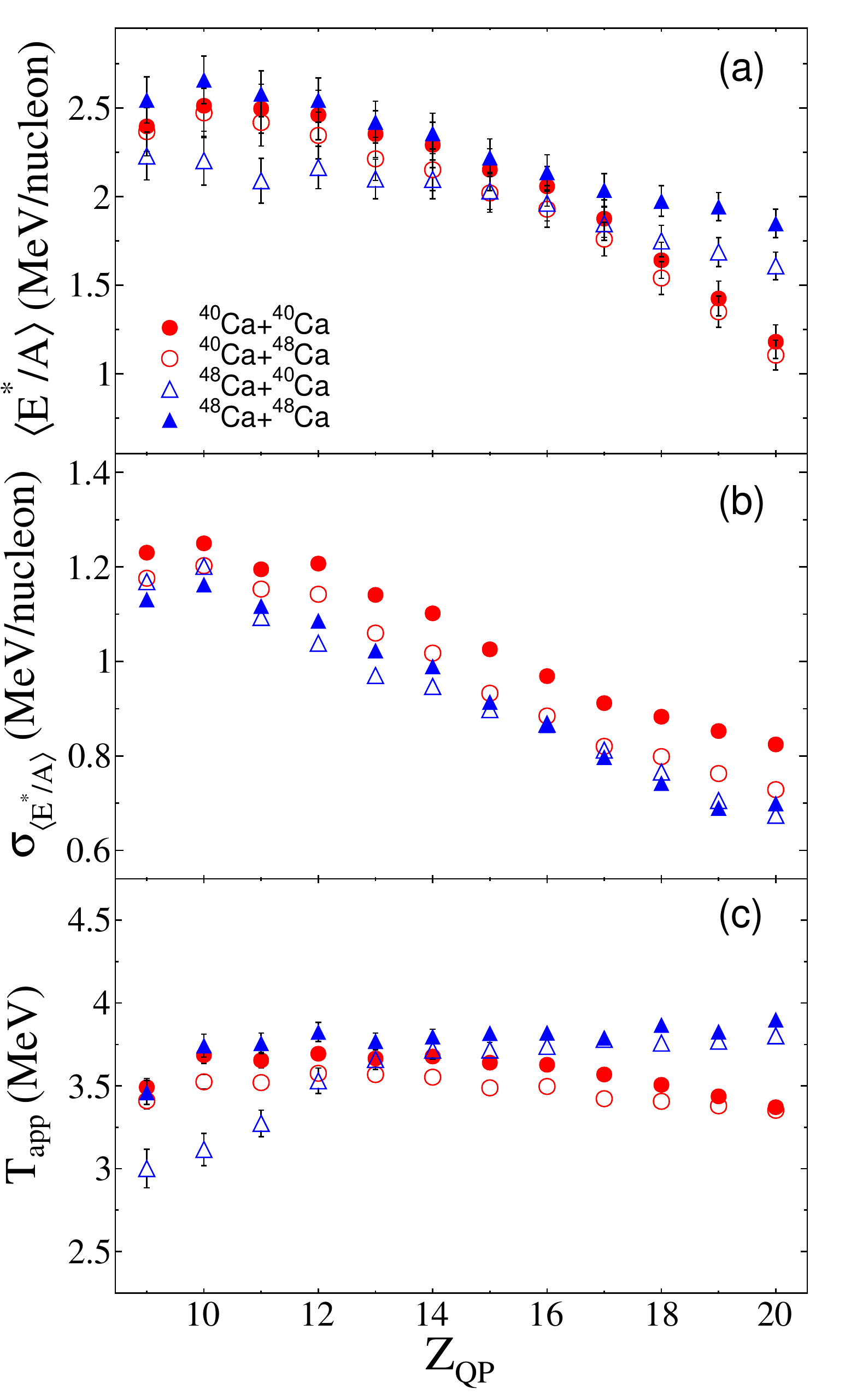}
\caption{(Color online) 
(a) Average excitation energy per nucleon of the reconstructed QP source as a function of its charge.
(b) Standard deviation of the excitation energy per nucleon distribution. 
(c) Apparent temperature, extracted from proton kinetic energy spectra (see text).
The values are averaged over $k=0.6-0.8$ (see text).}
\label{fig_EstarA_Tapp}
\end{figure}

Fig.\ref{fig_EstarA_Tapp}(a) depicts the evolution of the average excitation energy per nucleon of the reconstructed QP as a function of its charge.
We first observe a decreasing average excitation energy with increasing QP charge, with a minimum close to the charge of the projectile for all the systems. 
Minima around $1.25$ and $1.75$ MeV/nucleon are respectively obtained for the $^{40}$Ca and $^{48}$Ca respectively, while the maxima reach close to $2.5$ MeV/nucleon for all systems. 
The difference in the minima can be explained by the fact that the grazing angles for the n-rich projectiles are smaller than for the n-poor ones, therefore more dissipative reactions are triggered in the former case. This effect is also reproduced within the filtered model.
Furthermore, in the specific case of $^{48}$Ca+$^{40}$Ca reaction the lowest magnetic rigidity setting ($B\rho_0\simeq0.78$ $T\,m$) was not measured, compared to the other systems. 
This could explain the difference in maxima observed in Fig.\ref{fig_EstarA_Tapp}(a) for this system when $Z_{QP}<13$.

The evolution of the associated standard deviation is presented in Fig.\ref{fig_EstarA_Tapp}(b) where we observe increasing values with decreasing size of the QP, similarly to Fig.\ref{fig_EstarA_Tapp}(a).

It is also possible to determine experimentally the temperature of the QP source from the LCP measured in INDRA.
Nonetheless, depending on the thermometer method and the probe, the extraction of experimental temperatures may lead to an ordering of their values according to the LCP species \cite{PhysRevC_87_034617, McIntosh2014}.
Such variations may be related to the difference in the average emission time of the considered LCP and also to the difference in the average density of the source.
The precise characterization of the temperature fluctuations according to the chosen method and probes is out of the scope of the present work.

We only present in this Section the temperatures extracted from the slopes of the proton kinetic energy spectra in the reconstructed QP frame.
This method is nonetheless sensitive to contamination from particles at high energy in the QP frame, coming from the mid-rapidity region for example.
As demonstrated in \citep{Vient2018PhysRevC_98_044611}, the QP de-excitation can be characterized using a restricted spatial domain in order to select the protons (or other LCP) solely emitted by the QP.
The so-called ``3D Calorimetry$"$ method allows to compute the kinetic energy spectra of the identified LCP in the reconstructed QP frame, keeping only LCP emitted in a spatial domain where the QP acts as screen to other emission sources, such as pre-equilibrium emissions. 
We have applied this method and projected the selected proton velocities in the reaction plane, defined by the reconstructed QP velocity and the beam direction. 
In summary, the polar and azimuthal angles are respectively defined as (i) the angle between the vector normal to the reaction plane and the velocity vector of the LCP in the QP frame, (ii) the QP velocity vector in the center-of-mass frame and the normal projection of the velocity of the LCP on the reaction plane.
Six areas ($60^{\circ}$ wide) in azimuthal angles were then used to build the polar angular distributions of the selected LCP (in the reconstructed QP frame).
Similarly to \citep{Vient2018PhysRevC_98_044611}, we found that the forward domain presents isotropic distribution of the polar angle of the LCP, compatible with an evaporation of LCP from the QP de-excitation.

\begin{figure}[ht]
\centering
\includegraphics[scale=0.47]{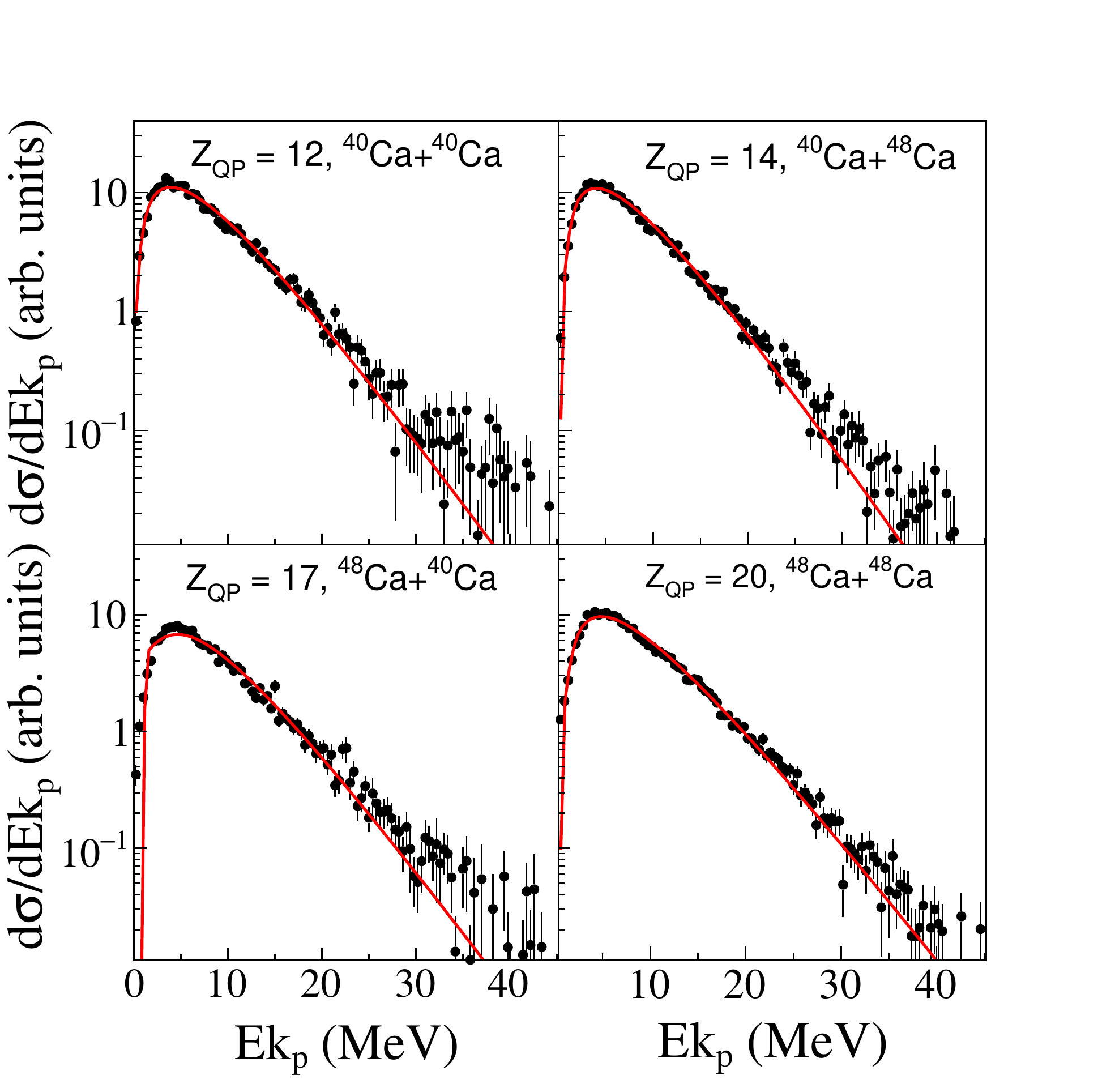} 
\caption{(Color online) Proton energy spectra (filled circles) fitted with a Maxwellian source (red lines) for various $Z_{QP}$ and systems, as indicated in the panels.}
\label{fig_temp_Zplf}
\end{figure}

The apparent temperatures were extracted by fitting the slope of the proton (selected for the QP reconstruction) kinetic energy spectra in the forward domains with a Maxwell-Boltzmann distribution \cite{PhysRev_52_295, Vient2018, BORDERIE201982}.
Some examples of representative spectra together with the Maxwellian fits, in the reconstructed QP rest frame, are shown in Fig.\ref{fig_temp_Zplf} for several primary charges, $Z_{QP}$= 12, 14, 17, and 20. The associated fits (red curves) provide a reasonable representation of the proton energy spectra.

The resulting apparent temperatures are presented in Fig.\ref{fig_EstarA_Tapp}(c).
We observe a matching of the distributions according to the neutron-richness of the projectile, with relatively stable temperature around $3.75$ MeV for the n-rich $^{48}$Ca projectile reactions, while the temperature increases from $3.25$ to $3.5$ MeV with decreasing $Z_{QP}$ for the $^{40}$Ca projectile reactions.
In addition to the aforementioned difference in the grazing angle, such pairing can also be explained by the use of proton spectra. 
Indeed, the decay of an excited projectile is expected to be preferentially via neutron emission for the neutron-rich projectile reactions. 
Thus proton emission is expected for more excited sources in the case of $^{48}$Ca projectile reactions compared to $^{40}$Ca.
Another interpretation could be the asymmetry dependence of the nuclear temperatures.
The overall values are nonetheless compatible with the compilation of \cite{natow01:T_limite}.
Finally, similarly to the excitation energy, the missing magnetic rigidity setting for the $^{48}$Ca+$^{40}$Ca reactions could explain the drop of temperatures for $Z_{QP}<13$ in Fig \ref{fig_EstarA_Tapp}(c).

For more consistency, those results were compared to (i) other LCP and (ii) the helium and hydrogen isotopes ratios \cite{Alb85, trautmann2007PhysRevC.76.064606}.
Concerning the extraction from other LCP kinetic energy spectra, similar trends are obtained from $^{2,3}$H and $^{3,4}$He. 
The $^{4}$He absolute values are close to the proton one with a difference of $10\%$ at worst, while higher values around $4.5$ MeV are obtained for deutons, tritons and $^{3}$He fits.
Concerning the second method, a difference of $8\%$ at worst is observed.

\begin{figure}[ht]
\centering
\includegraphics[scale=0.44]{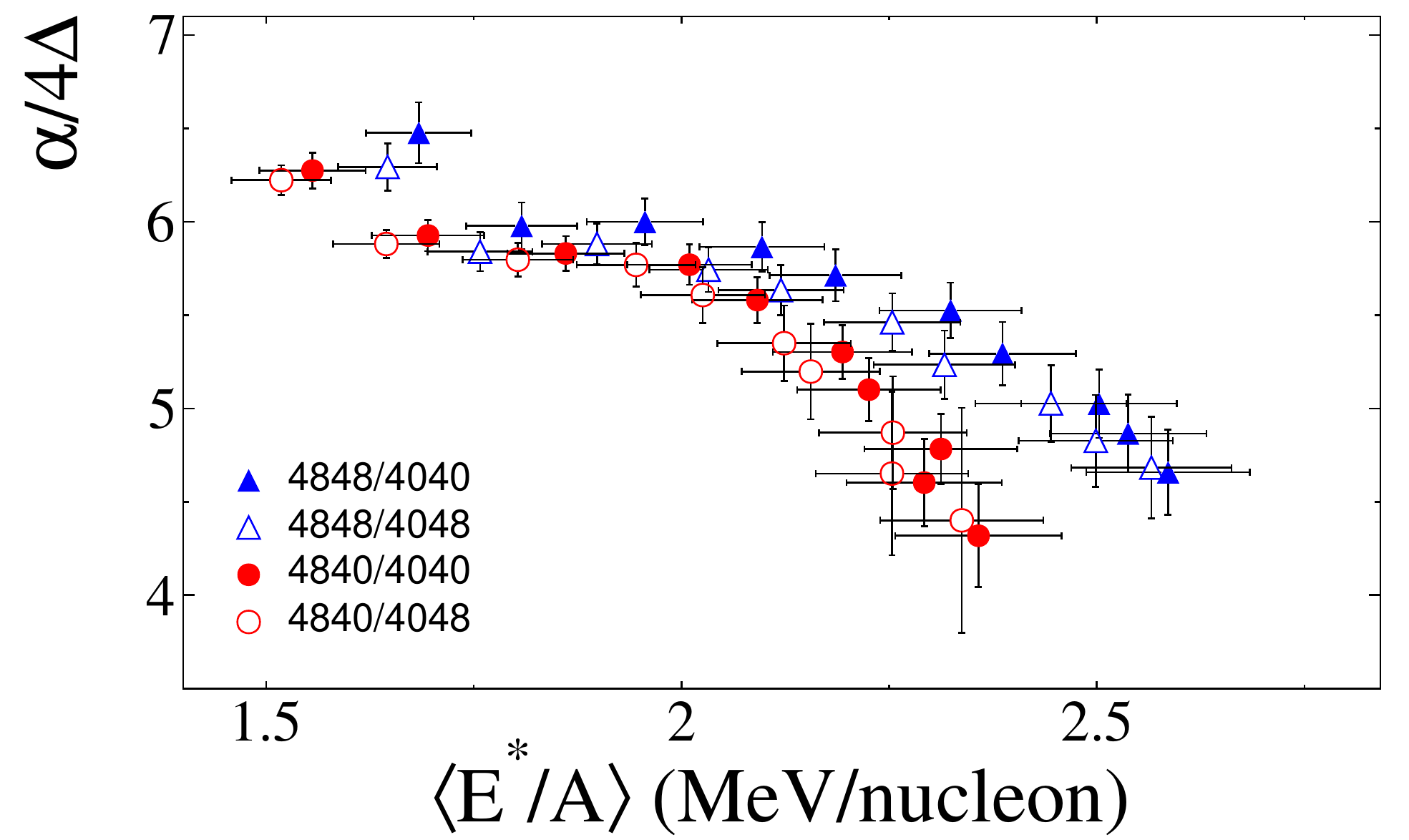}
\caption{(Color online) Experimental $\alpha/4\Delta = C_{sym}/T $ (see Eq.\ref{eq_csymalpha}) extracted from the isoscaling method as a function of the average excitation energy of the reconstructed QP source with the evaporated neutrons contribution.
The values are averaged over $k=0.6-0.8$ (see text).}
\label{fig_CsymT_EstarA}
\end{figure}

Fig.\ref{fig_CsymT_EstarA} depicts the behaviour of $\alpha/4\Delta$ as a function of the average excitation energy per nucleon of the reconstructed QP for $10 \leq Z_{QP} \leq 19$.
We observe, for all combinations, decreasing values of $\alpha/4\Delta$ as a function of increasing excitation energy.
This behaviour is consistent with various HIC isoscaling data \cite{shetty2007:densityDependenceEsym, PhysRevC_79_061602, lefevre2005_PhysRevLett_94_162701} and could be indicative of a decrease in symmetry energy as a function of increasing excitation energy. Comparisons of isoscaling multifragmentation data with evaporation models have also highlighted that $\alpha$, $C_{sym}$, the temperature of the source and the density at break-up are all correlated, and thus a drop in $\alpha/4\Delta$ may be related to a decrease in density \cite{shetty2007:densityDependenceEsym}.   

\subsection{Discussion concerning the symmetry energy}\label{subsec_discussion}

We would like to add some comments about the results presented in the previous sections and the extraction of relevant information on the symmetry energy term.

First, such a unique experiment allows to apply for the first time the isoscaling method directly to the reconstructed primary fragment for peripheral collisions.
In that regard, the results remain consistent within the grand-canonical hypothesis, where $C_{sym}$ reflects the symmetry energy of primary fragments at finite temperature \cite{PhysRevC_101_064314}.

Second, we want to underline that for such collisions the sequential decay of hot primary fragments have an effect on the experimental isoscaling parameters, proving the relevance of measuring the evaporated neutrons, which is an experimental challenge itself.

Third, we would like to stress the importance of measuring both $\alpha$ and $\Delta$ parameters to extract possible trust-worthy information on $C_{sym}$.
Indeed, depending on the system combination used in the isoscaling analysis, a different evolution of $\alpha$ and $\Delta$ as a function of the QP size can be observed.
This supports the idea that the behaviour of $\alpha$ itself is not sufficient to conclude about finite-size effects of the symmetry energy term.
Moreover, the extracted $\alpha / 4 \Delta$ ratios present an increase as a function of the QP charge, almost equivalent for all system combinations, which could be interpreted as an experimental evidence of a strong surface dependence. 
Even with the drawbacks of the evaporated neutrons estimation used in this analysis, we think that these results remain consistent as a better linearity of the $\alpha$ and $\Delta$ parameters is observed.

Finally, we must highlight the caveats concerning the extraction of the experimental nuclear temperatures.
The saturation of $\langle E^{*}/A \rangle$ and $\sigma_{E^{*}/A}$ observed in Fig.\ref{fig_EstarA_Tapp} for $Z_{QP} < 14$ proves that the sensitivity to the most dissipative collisions is reduced for small QP size for the reactions under study.
Thus, the selection on $Z_{QP}$, while necessary for isoscaling, is not restrictive enough for extracting the temperature, independently of the thermometer method (and probe). 
In fact, the results presented in Section \ref{subsec_calo} reflect an average value of various excited sources having a wide excitation energy distribution. 
Furthermore, assuming the validity of a thermometer, one needs also to consider several corrections of the extracted values, such as finite-size, emission time differences or secondary decay effects \cite{Kelic2006}.
At the present stage of this work, the data are clearly not well suited to deduce $C_{sym}$ from the apparent temperatures. 
As a first step to extract relevant information on $C_{sym}$, one should at least consider to apply a selection on the excitation energy and correct the apparent temperatures of the de-excitation cascade of a given source, by the help of statistical model simulations.

Nonetheless, considering the previous comments, we think that the present analysis shows promising results concerning the possibility to extract the surface-to-volume contribution of $C_{sym}$ from isoscaling.
In this direction, we also plan to compare the isoscaling results to other approximations of the symmetry energy term, based on the direct use of the widths of the isotopic distributions \cite{PhysRevC_75_024605, ono2004:SymmEnergy}.

\section{Conclusion\label{sec:Conclusion}}

In this work, an experimental study of semi-peripheral to peripheral $^{40,48}$Ca$+^{40,48}$Ca collisions at $35$ MeV/nucleon was presented.
The experimental set-up consisted of the VAMOS high acceptance spectrometer and the INDRA $4\pi$ multi-detector array at GANIL. 
VAMOS was positioned at forward angles ($2.5^{\circ}$-$6.5^{\circ}$) to measure the PLF (QP remnant) in its focal plane with a high isotopic resolution of the order of $\Delta A/A \approx 0.7\%$, in coincidence with the charged particles detected in INDRA covering the remaining (almost) 4$\pi$ solid angle.
The collected data present a good detection efficiency and correct correlations between the two devices, allowing coincident measurements of the QP residue and emitted charged particles.

The isotopic composition of the forward-emitted PLF exhibits a neutron enrichment according to the neutron richness of the projectile but also, to a lesser extent, of the target.
Furthermore, a similar behaviour is observed for the variance of the isotopic distributions, which can be related to the isoscaling observable \cite{Raduta2007, Ono_PRC_68_051601, PhysRevC_70_011601}. 

Comparing the average multiplicity distributions of the forward emitted light charged particles as a function of the PLF size, we also observed trends according to the neutron enrichment. 
Indeed, neutron-rich $t$ and $^{6}He$ (respectively protons and neutron-deficient $^{3}He$) particles demonstrate a clear hierarchy of the multiplicities according to the neutron-richness (respectively neutron-deficiency) of the projectile and, to a lesser extent, of the target.
The aforementioned results can be interpreted as an experimental evidence of the isospin diffusion mechanism.

A reconstruction method of the primary fragment is proposed, by associating event-by-event the PLF with the identified and selected particles detected with INDRA. 
In order to exclude pre-equilibrium and neck emissions, a selection of the LCP based on their correlations with the PLF and the TLF was applied. 
Numerical cuts were defined, based on their relative velocities so as to keep only particles emitted by the QP, in agreement with the model calculations. 
Also, as the neutrons evaporated by the QP were not measured, the mean neutron multiplicities extracted from AMD followed by GEMINI++ were used as a surrogate.

A study of the isoscaling method was conducted, based on three different species: the PLF and the reconstructed QP with or without the evaporated neutrons estimation.
The isoscaling was observed from the yield ratios of the PLF and the reconstructed QP, for all system combinations.
A linear correlation between the extracted isoscaling $\alpha$ parameter and the average neutron composition of the two sources $\Delta$ is only observed for the reconstructed QP, leading to the conclusion that the reconstruction of the PLF is mandatory for the reactions under study.
A clear evolution of both parameters is also observed as a function of the size of the reconstructed QP, the latter directly correlated to the excitation energy and the temperature as it reflects the centrality of the collision.  
This could be interpreted as an experimental evidence of a strong surface dependence of the symmetry energy term.
Furthermore, the same hierarchy is observed for both parameters, according to the neutron-richness of the system combination used to apply the isoscaling method.
This hierarchy is an experimental evidence that justifies the use of the $\alpha$ parameter as a surrogate for isospin asymmetry in isospin transport studies.   

The reconstruction of the QP allowed to estimate its excitation energy using calorimetry method. 
For the most dissipative collisions, the average excitation energies are about $2.5$ MeV/nucleon, while they range from $1.25$ and $1.75$ MeV/nucleon for the $^{40}$Ca and $^{48}$Ca respectively. 
This difference is attributed to the difference in grazing angle of the systems.
Applying a ``3D Calorimetry$"$ method, the apparent temperatures were also extracted from the slope of the proton kinetic energy spectra, leading to relatively stable values in the $3.25-3.75$ MeV range for all systems.
A decreasing $\alpha/4\Delta$ with increasing excitation energy is observed, in agreement with existing results obtained with similar methods \cite{lefevre2005_PhysRevLett_94_162701, SOULIOTIS200435, PhysRevC_73_024606, PhysRevC_75_011601}.
According to the standard isoscaling formula, $\alpha/4\Delta = C_{sym}/T$, this behaviour could be related to the change of the symmetry energy term.  

As a conclusion, we think that the experimental results presented in this paper bring more information about the suitability and the limits of the isoscaling method in peripheral collisions at intermediate energies. 
We have highlighted that the experimental $\alpha$ and $\Delta$ parameters are distorted due to secondary decays, but also present a dependence on the considered system combination that could justify the use of $\alpha$ as a surrogate for isospin asymmetry in isospin transport studies.
The reconstruction of the QP is mandatory to observe an evolution of $\alpha/4\Delta$ with the size of the QP, the latter being consistent with an effect of strong surface contributions to the symmetry energy term in finite nuclei.  
Moreover, we would like to stress that the experimentally determined temperatures are additional sources of uncertainties as they are estimated directly from the measured fragments, making a direct estimation of the symmetry energy term from isoscaling difficult for the present analysis. 
The present data show indeed slight differences in the apparent temperatures according to the system, which can be explained by the extraction of the temperatures from proton kinetic energy spectra.
  
Finally, the isotopic composition of the PLF and the reconstructed QP, along with the associated LCP multiplicities present promising information for the study of isospin diffusion. 
This work is currently in progress.



\begin{acknowledgments}
The INDRA collaboration would like to dedicate this article to Marie-France Rivet and Elio Rosato, both of whom have sadly deceased since the experiment was performed in 2007. Their contributions not only to this experiment but to the scientific advancements of the INDRA collaboration since its beginnings will not be forgotten.
The authors would like to thank:  
The staff of the GANIL Accelerator facility for their continued support during the experiments; A. Navin for his constant support;  M. Rejmund for setting up the VAMOS spectrometer, without his help the experiment could not succeed;
B. Lommel and the Target Laboratory of the GSI Helmholtzzentrum for providing the $^{48}$Ca targets; 
The Target Laboratory of Legnaro for providing the $^{40}$Ca targets;
Wilton Catford for providing TIARA electronics used for the CsI wall in the VAMOS focal plane; 
A. Lemasson and B. Jacquot for their invaluable help with VAMOS trajectory reconstructions.
The authors acknowledge the participation of P. St-Onge in the normalization procedure and in the calibration of the VAMOS spectrometer.
A. C. and Q. F. acknowledge productive discussions with A. Ono, W. Trautmann and S. Typel.
Q. F. gratefully acknowledges the support from CNRS-IN2P3 and R\'egion Normandie (France) under RIN/FIDNEOS.
\end{acknowledgments}

\appendix

\section{Fragment identification with VAMOS}\label{app1_VamosID}

\subsection{Trajectory reconstruction}\label{app11_event_recon}

VAMOS is a software spectrometer in the sense that trajectory reconstruction technique must be used to determine the momentum and scattering angles of the particles in the laboratory frame, from the measured quantities in the focal plane. Indeed, the large acceptance of the spectrometer induces significant non-linearities that can only be estimated from transfer map calculations. 
The trajectory reconstruction aims to deduce the initial parameters $(\theta, \phi, B\rho, L)$ from the final positions $(x_f,y_f)$ and angles $(\theta_f, \phi_f)$ measured at the focal plane, where $(\theta, \phi)$ are the scattering angles of the particle in the laboratory frame, $B\rho$ the magnetic rigidity and $L$ the path length from the target to the stopping detector. 

A precise computation of trajectories was done by simulating the ion trajectories through the spectrometer using the ion optical ray-tracing code \textsc{zgoubi} \cite{MEOT1999353,MEOT2014112}.
This code allows the tracking of arbitrarily large numbers of ions through a given set of optical elements. In summary, the input of \textsc{zgoubi} are the geometry and relative location of the optical components of the VAMOS line used in the present experiment, the associated field maps and finally a large set of ions to be transmitted in the previously defined VAMOS geometry. 
The particles are described by three parameters $(\delta,\theta,\phi)$, where $\delta=B\rho/B\rho_{0}$ is the relative magnetic rigidity, $\theta$ is the angle between the $z$-axis and the projection of the velocity vector of the particle on the $xz$ plane and $\phi$ is the angle between the velocity vector and its projection on the $xz$ plane.

In order to reduce the complexity of the usual high-order polynomial calculations used to reconstruct the trajectories, we have developed a method similar to the one described in \cite{Pullanhiotan2008343:VAMOS, Ramos_2018:PhysRevC.97.054612}. 
A dataset of trajectories, covering the full acceptance of the spectrometer, was computed from \textsc{zgoubi} and decomposed into small bins in the final coordinates plane $(x_f,y_f,\theta_f,\phi_f)$.
The reconstruction of the trajectory parameters was then applied by selecting a subset of trajectories close to the region of interest, using the minimum square distance to the experimental coordinates at the focal plane. A local polynomial fit was finally applied to extract the trajectory from the subset \cite{fable:tel-01775269}.

\subsection{Particle identification}\label{app12_vamos_id}

The following procedure was applied to identify the fragments detected in VAMOS :

i) Reconstruction of the particle position and scattering angles at the focal plane ($x_{f},y_{f}, \theta_{f},\phi_{f}$) from the positions measured in the two drift chambers ;

ii) Reconstruction of the emission angles at the target position ($\theta, \phi$), the magnetic rigidity $B\rho$ and the length path $L$ of the fragment using the four previous parameters and \textsc{zgoubi} simulations ;

iii) Identification of the atomic number, $Z$, utilizing the $\Delta E-E$ identification, from either ionization chamber-Si or Si-CsI telescopes ; 

iv) Determination of the time-of-flight ($ToF$) of the particle ;

v) Determination of the mass number to charge state ratio, $A/Q$, and the mass number from the measured energy losses $A_{E}$, using the two relationships:
$A/Q=\frac{B\rho}{3.107\beta\gamma}$ and $A_{E}=\frac{2E}{931.5\beta^{2}}$ where $B\rho$ is the magnetic rigidity of the fragment in $T\,m$, $\beta=v/c$ its reduced velocity (deduced from the $ToF$) and $E$ its measured total kinetic energy.

vi) Identification of the charge state $Q$ from the 2-dimensional map of $A_{E}$ versus $A/Q$. A calculated grid was drawn in this map and adjusted in order to assign a charge identification (CID) such as $CID=Q+\delta Q$, where $\delta Q$ is the distance to the closest line in the grid. 

vii) Determination of the reconstructed mass number from the integer value of the CID : $A_{V}=\lfloor CID \rfloor \cdot (A/Q)$.
\newline

\begin{figure}[ht]
\centering
\includegraphics[scale=0.7]{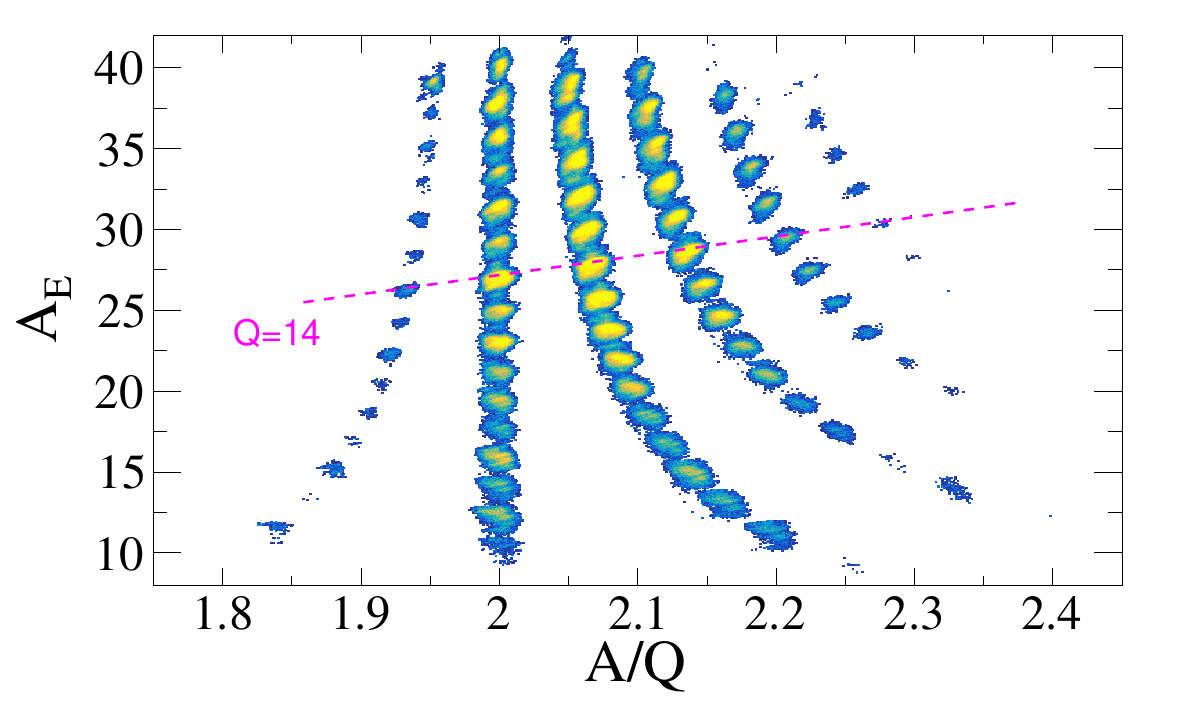}
\caption{(Color online) Mass number ($A_{E}$) as a function of mass number over the charge state ($A/Q$) of the detected fragments for the $^{40}Ca+^{48}Ca$ reaction. The dotted line corresponds to charge state $Q=14$.}
\label{fig_ae_aoq}
\end{figure}

\begin{figure}[ht]
\centering
\includegraphics[scale=0.4]{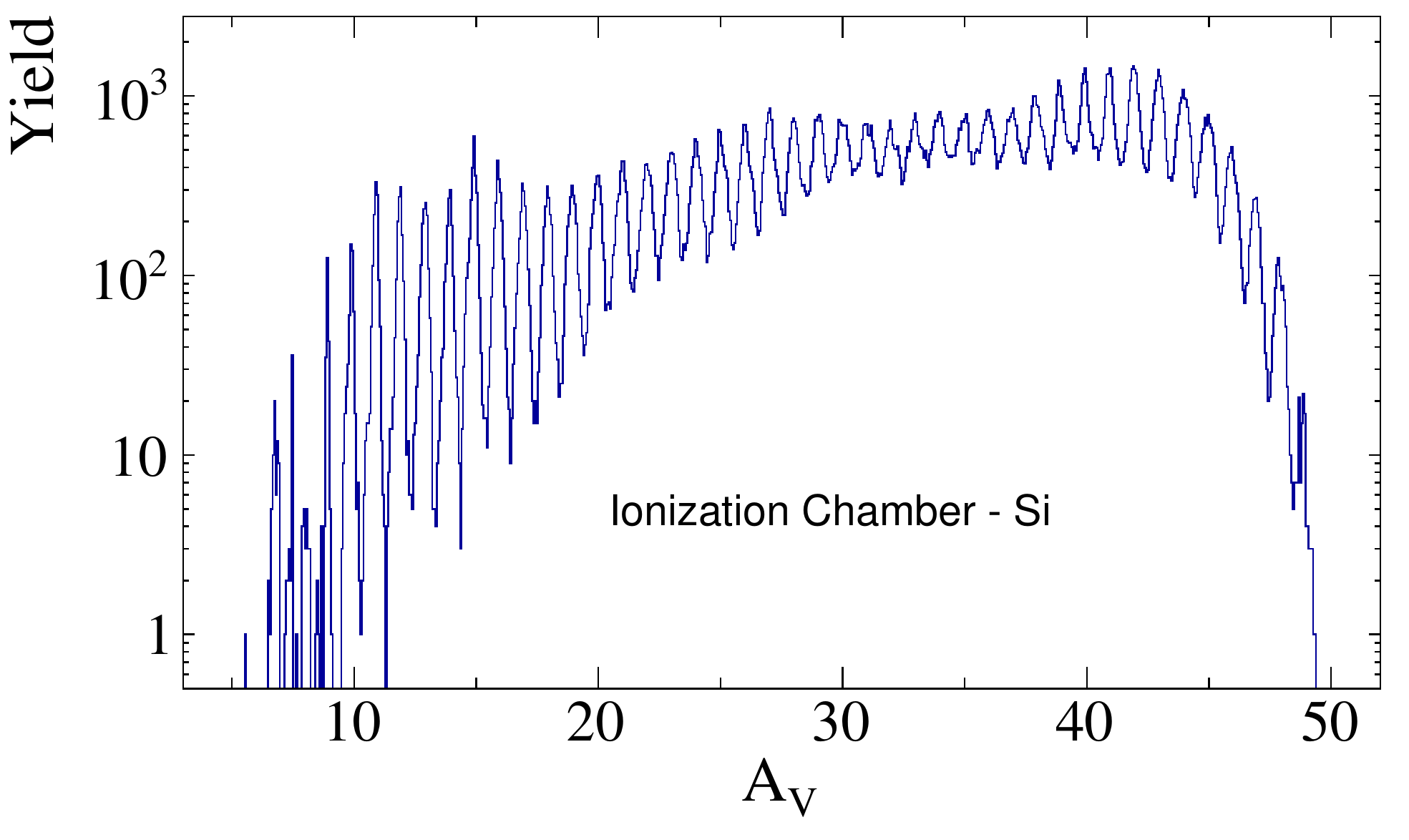}
\includegraphics[scale=0.4]{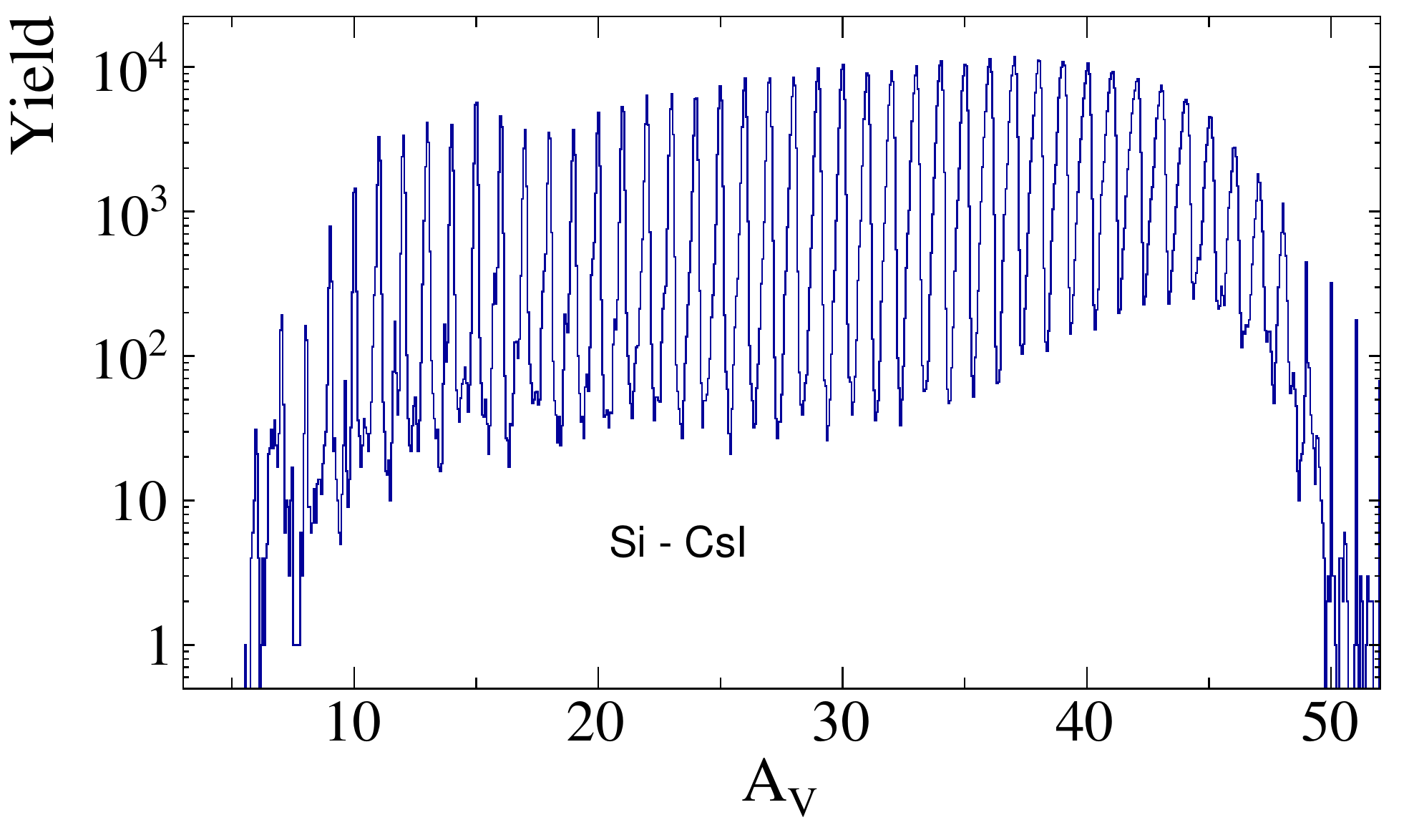}
\caption{(Color online) Mass number of the fragments detected in Ionization Chamber-Si (upper panel) and Si-CsI (lower panel) telescopes for the system $^{48}Ca+^{48}Ca$.} 
\label{realA}
\end{figure}

Fig.\ref{fig_ae_aoq} shows the mass number $A_{E}$ vs the mass number over the charge state $A/Q$ of the fragments measured in the $^{40}$Ca+$^{48}$Ca reaction. A correction of the order of 1 $ns$ of the ToF, based on $A/Q = 2$ alignment, was necessary to obtain this sharp identification \cite{fable:tel-01775269}.
Fig.\ref{realA} shows the reconstructed mass distributions for fragments detected in the Si-CsI and ionization chamber-Si telescopes in the VAMOS focal plane for the $^{40}Ca+^{48}Ca$ reaction.    
A mass resolution of about $\Delta M/M\sim1/285$ has been obtained for $A=40$.

More details about the trajectory reconstruction and the VAMOS performances can be found in \cite{Pullanhiotan2008343:VAMOS} and a detailed description for the present experiment is given in \cite{fable:tel-01775269}.

\section{Normalization of the events}\label{app2_normalization}

In order to normalize the statistical weight of the events, the overlap between magnetic rigidity settings needs to be considered, along with the acceptance of the spectrometer, the variations in beam intensity and dead time of the acquisition system. 
In this section we describe the normalization procedure applied to the measured events in order to correctly reproduce the reaction kinematics without biasing the analysis \cite{fable:tel-01775269}.

\subsection{Acceptance of the spectrometer}\label{app21_vamos_acceptance}

The ion trajectory in the spectrometer depends not only on the reaction kinematics but also on the acceptance of the spectrometer. The former defines the momentum distribution and angles of the ions entering the spectrometer, while the latter limits the range of ion momenta and angles reaching the focal plane \cite{Ramos_2018:PhysRevC.97.054612}. Indeed, setting the magnetic rigidity of the spectrometer at a nominal value $B\rho_{0}$ limits the range of $B\rho$ values for accepted trajectories around that value. We have estimated this range to be $10\%$ of the nominal value. Steps of $8\%$ were then chosen in order to have an overlap between successive $B\rho_{0}$ settings.  

A detailed study of the acceptance of VAMOS, inspired from \cite{Pullanhiotan2008343:VAMOS}, was carried out by simulating the ion trajectories through the spectrometer using the ion optical ray-tracing code \textsc{zgoubi} \cite{MEOT1999353,MEOT2014112,fable:tel-01775269}.

In order to cover the whole experimental acceptance of VAMOS, more than $10^{7}$ trajectories were computed with an angular step adapted to the experimental resolution (exceeding the experimental aperture of the spectrometer). The geometry of the detection chamber and the $4.5^{\circ}$ rotation of the spectrometer with respect to the beam direction were also taken into. 
To validate the geometry used in the \textsc{zgoubi} simulation, the focal plane observables from the simulation were directly compared with the one measured from the experiment. The latter was completely covered by the simulated focal plane. Finally, cuts on $(\theta, \phi)$ were applied based on the experiment, so as to eliminate the aberrations in positions and angles. A good agreement between the data and simulation has been obtained.

As stated in \cite{Pullanhiotan2008343:VAMOS}, the measured counts in VAMOS is influenced by the variation of the solid angle as a function of rigidity, whereas no functional formula is able to describe it. The proposed correction in acceptance is to separate the data into finite bins in $\delta$ and $\theta$: $\delta \in \left[ \delta - \Delta \delta, \delta + \Delta \delta \right[$ and $\theta \in \left[ \theta - \Delta \theta,  \theta + \Delta \theta \right[$. We were then able to define an effective solid angle such as:
\begin{equation}\label{eq_diff_solidangle}
\Delta ^2 \Omega (\delta, \theta) = \int _{\theta - \Delta \theta}^{\theta + \Delta \theta} sin \theta \hspace{0.1cm} d\theta \int_{\phi^{min}(\delta, \theta)}^{\phi^{max}(\delta, \theta)}d \phi
\end{equation}
where the integration is performed over the bin size of $\theta$ and the $\phi$ acceptance limits $(\phi_{min}, \phi_{max})$ given by \textsc{zgoubi}.  

By definition, this effective solid angle can also be related to the geometrical efficiency $\epsilon_{geo}$ of the spectrometer in the $(\delta, \theta)$ domain :
\begin{equation}
\epsilon_{geo}(\delta,\theta) = \frac{\Delta^{2}\Omega(\delta,\theta)}{4\pi} 
\label{eq_epsilon_geo} 
\end{equation}

\subsection{Beam intensity}\label{app22_beam}

In this experiment, the overall trigger was an ``OR$"$ of  VAMOS Si detectors, thus only INDRA triggers validated by VAMOS were retained by the data acquisition system.
Since the rate at which reactions are produced by the beam particles impinging on the target is independent of the $B\rho_{0}$ setting of VAMOS, and as the INDRA trigger rate is roughly proportional to the reaction rate, we have used the recorded scaler values for the INDRA trigger as a measure of beam intensity. This scaler was recorded throughout the experiment, whether or not the event was accepted by the acquisition. 
We also checked in a separate direct intensity measurement with a Faraday Cup that the scaler of the INDRA trigger is proportional to the beam intensity (when acquisition dead time is taken into account). 

\subsection{Normalization procedure}\label{app23_normalization}

For each system, we aim to apply a weight $W$ including all corrections on an event by event basis, so that the sum of the detected particle distributions reflects the actual particle distribution of the collision, thus :
\begin{equation}
W(B\rho,\theta).\sum_{j} Y^{j}_{D}(B\rho,\theta)  \propto  Y^{tot}_{R}(B\rho,\theta)
\label{eq_w1} 
\end{equation}
where $j$ defines a specific nominal magnetic rigidity setting $B\rho_0^{j}$, $Y^{j}_{D}(B\rho,\theta)$ and $Y^{tot}_{R}(B\rho,\theta)$ are respectively the measured counts of particles for the $B\rho_0^{j}$ setting and the total amount of emitted particles over all magnetic rigidity settings, for a given $(B\rho, \theta)$ domain.

By definition, the number of detected particles for a given $B\rho_0^{j}$ setting depends on the total amount $Y^{j}_{R}(B\rho,\theta)$ of emitted particles in the $(B\rho = \delta \cdot B\rho_0^{j}, \theta)$, the detection efficiency and the associated dead time, such as:
\begin{equation}
 Y^{j}_{D}(B\rho,\theta)=Y^{j}_{R}(B\rho,\theta)\epsilon_{int}\epsilon^{j}_{geo}(B\rho,\theta)(1-DT^{j})
\label{eq_w2} 
\end{equation} 
where $\epsilon_{int}$ is the intrinsic detection efficiency (supposed to be constant), $\epsilon^{j}_{geo}$ the geometrical efficiency defined in Eq.\ref{eq_epsilon_geo}, and $DT^{j}$ the dead time of the acquisition system for the $B\rho_0^{j}$ setting.
 
These previous distributions can also be related to the reaction cross section by:
\begin{equation}
Y^{tot}_{R}(B\rho,\theta)=\sigma_{R}.N_{inc}^{tot}
\label{eq_sig1} 
\end{equation}
\begin{equation}
Y^{j}_{R}(B\rho,\theta)=\sigma_{R}.N_{inc}^{j}
\label{eq_sig2}  
\end{equation}
where $N_{inc}^{j}$ is the amount of incident particles for the given $B\rho_0^{j}$ setting and $N_{inc}^{tot}$ is the total amount of incident particles reaching the target over all nominal magnetic rigidity settings, respectively.

\begin{figure}
\centering
\includegraphics[scale=0.27]{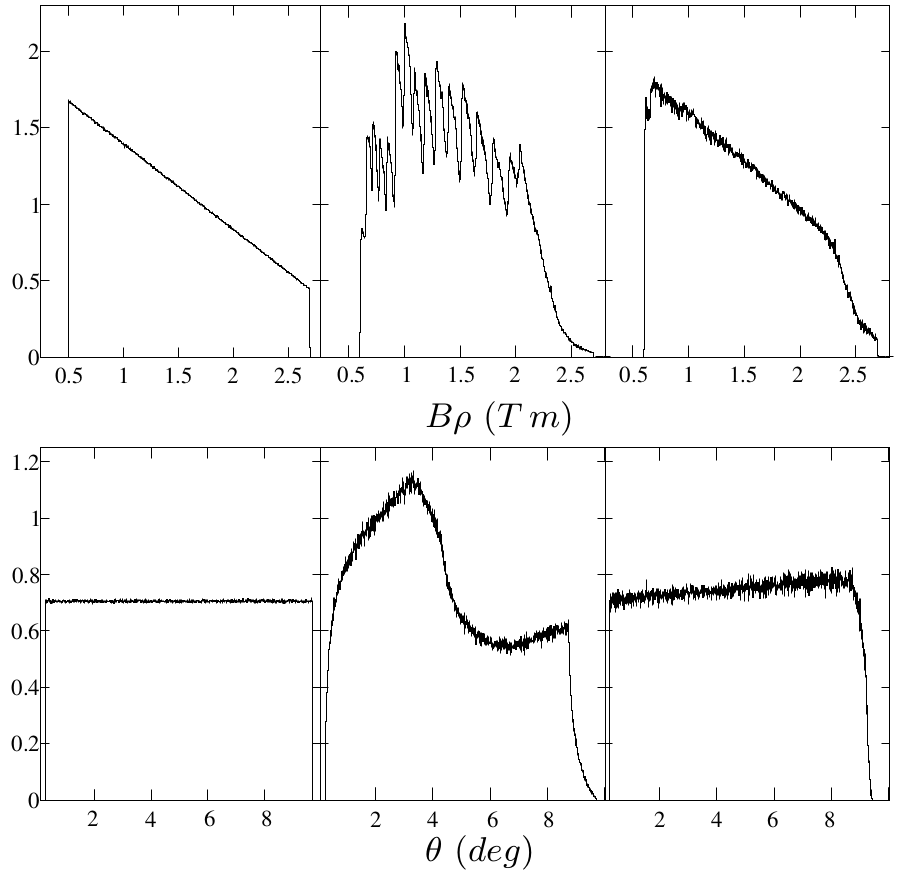}
\caption{Example of the performed Monte Carlo simulations showing the effect of the normalization. A triangular distribution of magnetic rigidity and uniform distribution of polar angle of the emitted particles were used to simulate a random number of trajectories in \textsc{zgoubi} up to VAMOS Si-wall, over all experimental nominal magnetic rigidity settings. Left, center, right hand panels represent respectively the input distributions that should be reproduced, the sum of the distributions obtained after the simulation and the distribution after the normalization.}
\label{fig_ex_norm}
\end{figure}

%
%

Therefore, according to the assumption that the trigger rate of INDRA is proportional to the number of incident ions and if $S_{j}$ is the recorded scaler values for INDRA for given $B\rho_0^j$ setting, one can deduce from Eqs.\ref{eq_sig1} and \ref{eq_sig2} :
\begin{equation}
Y^{j}_{R}(B\rho,\theta) = Y^{tot}_{R}(B\rho,\theta)\frac{S_{j}}{\sum_{j'} S_{j'}} 
\label{eq_w3} 
\end{equation}

Finally, by combining Eq.\ref{eq_epsilon_geo}, \ref{eq_w1}, \ref{eq_w2} and \ref{eq_w3}, one obtains the following weight to be applied for each event:
\begin{equation}
W(B\rho,\theta) = \frac{C \cdot \sum_{j} S_{j}}{\sum_{j'}S_{j'} \epsilon_{int} \epsilon^{j'}_{geo}(B\rho,\theta)(1-DT^{j'})}
\label{eq_weight_event}
\end{equation}
where $C$ is a constant which depends on the reaction total cross section.

In order to check the aforementioned normalization method, Monte Carlo simulations based on \textsc{zgoubi} have been studied. Arbitrary number of particles were simulated up to VAMOS Si-Wall and CsI-wall, for each experimental nominal rigidity setting, following random $B\rho$, $\theta$, $\phi$ and dead time input distributions. 

Fig.\ref{fig_ex_norm} shows an example with a triangular distribution of magnetic rigidity and uniform distributions of polar and azimuthal angles. The leftmost figures represent the input distributions that are aimed to be reproduced, the central panels represent the overall output distributions after the simulation (representative of the experiment) and the right panels represent the distribution after the normalization. One can clearly see that the input distributions are better reproduced after the normalization. 

\section{Isoscaling fits}\label{app3_isofits}

The Gaussian approximation allows to express the ratio of the yields of the two systems (1) and (2) for a given isotope (fixed $Z$) such as \cite{Raduta2007}:
\begin{align}
\label{eq_iso_quad}
ln \left( \frac{Y_{(2)}(N,Z)}{Y_{(1)}(N,Z)} \right) = & -\frac{N^2}{2} \left( \frac{1}{\sigma^2_{N_{2}}} - \frac{1}{\sigma^2_{N_{1}}} \right) \\
 & + N \left( \frac{\langle N_{2} \rangle }{\sigma^2_{N_{2}}} - \frac{\langle N_{1} \rangle}{\sigma^2_{N_{1}}} \right)  +  K(Z) \nonumber
\end{align}

where $\langle N_{i} \rangle$ and $\sigma_{N_{i}}$ are the mean neutron number and deviation associated with element $Z$ for reaction $(i)$.

Furthermore, the isoscaling phenomenon observed in a variety of HIC is expressed as a linear dependence in $N$ at fixed $Z$, such as (see also Eq.\ref{eq_iso_params}):
\begin{equation}
\label{eq_iso_params2}
ln \left( \frac{Y_{(2)}(N,Z)}{Y_{(1)}(N,Z)} \right) = \alpha(Z) N + K(Z) 
\end{equation}

As a consequence, isoscaling is expected in a mass region where the Gaussian approximation is well verified and if $ \sigma^2_{N_{1}} \simeq \sigma^2_{N_{2}}$, so that Eq.\ref{eq_iso_quad} and \ref{eq_iso_params2} are equivalent.  
The aforementioned comment is of particular interest for the present analysis as a broad range of isotopes is measured with VAMOS, as can be seen in Fig.\ref{fig_nuchar}.

\begin{figure}[H]
\includegraphics[scale=0.5]{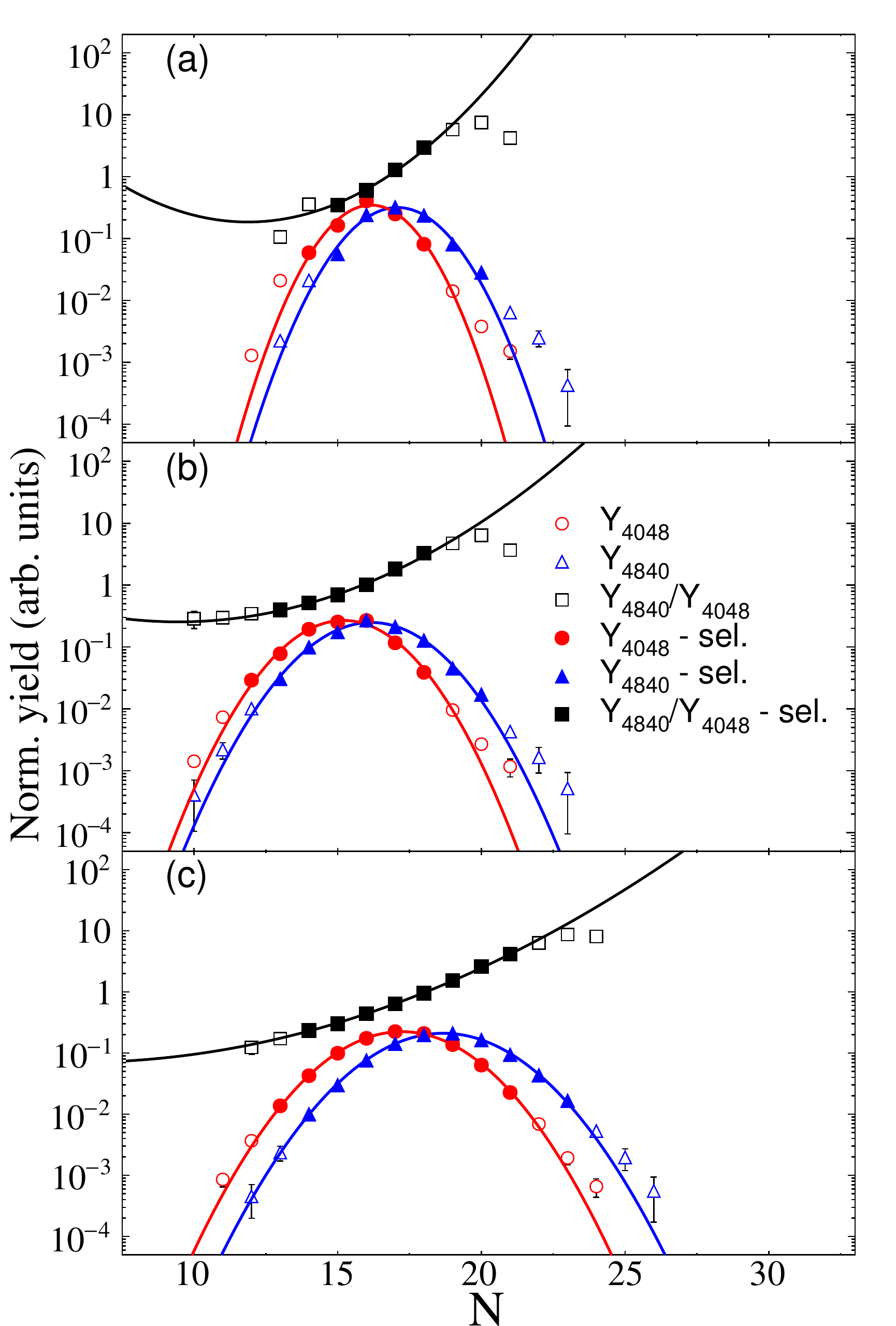}
\caption{(Color online) Illustration of the gaussian approximation with $Z=15$ isotopes, with the $^{40}$Ca+$^{48}$Ca and $^{48}$Ca+$^{40}$Ca systems combination, for (a) the fragment identified in VAMOS and the reconstructed QP (b) without or (c) with the evorated neutron contribution (see text).}
\label{fig_iso_test}
\end{figure}

Figure \ref{fig_iso_test} illustrates the quality of the Gaussian approximation (solid lines) for the experimental $N$ distributions (normalized to their integral) obtained for $Z=15$ isotopes, with the $^{40}$Ca+$^{48}$Ca (open circles) and $^{48}$Ca+$^{40}$Ca (open triangles) systems combination.

The ratio of the individual gaussian fits is represented in solid black lines while the experimental yields ratio (see Eq.\ref{eq_iso_params2}) is represented by open squares.
It can be seen that the gaussian approximation holds in a limited region of $N$. 
Thus, the isoscaling fit procedure was applied in a limited region of experimental points, within an empirical limit of three standard deviations of the mean of the individual Gaussian fits, represented by full symbols in Fig.\ref{fig_iso_test}.

\bibliographystyle{apsrev}
\addcontentsline{toc}{section}{\refname}
\bibliography{biblio}

\end{document}